\documentclass[longbibliography,prl,aps,twocolumn,showpacs,floatfix]{revtex4-2}
\usepackage{amsfonts}
\usepackage{amssymb}
\usepackage{hyperref}
\usepackage{graphicx}
\usepackage{dcolumn}
\usepackage{bm,amsmath,verbatim}
\usepackage{mathrsfs}
\usepackage{color}
\usepackage{lineno}
\usepackage{dsfont}
\usepackage{amsmath,amssymb,amsthm}

\usepackage[T1]{fontenc}
\usepackage[latin9]{inputenc}
\usepackage{booktabs}

\hypersetup{colorlinks,
	linkcolor=blue,          citecolor=blue,        filecolor=blue,      urlcolor=blue           }

\def\be{\begin{equation}}
	\def\ee{\end{equation}}
\def\bea{\begin{eqnarray}}
	\def\eea{\end{eqnarray}}
\def\bse{\begin{subequations}}
	\def\ese{\end{subequations}}

\def\be{\begin{eqnarray}}
	\def\ee{\end{eqnarray}}

\begin{document}
%\setlength\columnsep{25pt}
%\linenumbers
\title{Quantized Thouless Pumping of Dark Solitons}
\author{Yu-Liang Tao$^{1}$}
\thanks{These authors contributed equally to this work}
\author{Huaxin He$^{2}$}
\thanks{These authors contributed equally to this work}
\author{Hao Lyu$^{2}$}
\author{Yongping Zhang$^{2}$}
\email{yongping11@t.shu.edu.cn}
\author{Yong Xu$^{1,3}$}
\email{yongxuphy@tsinghua.edu.cn}
\affiliation{$^{1}$Center for Quantum Information, IIIS, Tsinghua University, Beijing 100084, People's Republic of China}
\affiliation{$^{2}$Institute for Quantum Science and Technology, Department of Physics, Shanghai University, Shanghai 200444, People's Republic of China}
\affiliation{$^{3}$Hefei National Laboratory, Hefei 230088, People's Republic of China}

\begin{abstract}
	Nonlinearity enables the emergence of localized waves such as solitons that maintain their shapes during propagation.
	Solitons are broadly classified into bright and dark solitons.
	While a bright soliton exhibits a density peak, a dark soliton presents as a defect on a continuous wave background. 
	A distinctive feature of dark solitons is the abrupt phase change in their wave function, which can host Majorana zero modes in
	topological fermionic superfluids. Recent studies have shown that bright solitons can undergo quantized transport through 
	Thouless pumping, where the bright soliton functions as a Wannier function. However, it remains unclear 
	whether Thouless pumping can also occur for dark solitons, which fundamentally differ from bright solitons. 
	Here, we theoretically demonstrate the occurrence of both integer and fractional Thouless pumping for  
	dark solitons within both a continuous model under optical lattices and a tight-binding model. Specifically, we find 
	that a dark soliton is transported by one or half a unit cell, following the center-of-mass position 
	of a Wannier function, as a system parameter is slowly varied over one cycle. 
	Our work opens new avenues for exploring Thouless pumping 
	for defects with phase changes, such as dark solitons, vortex solitons, ring dark solitons, and vortices.
\end{abstract}

\maketitle
Nonlinearity is prevalent across various disciplines, ranging from physics to social sciences. 
In physics, it has been extensively studied in photonic systems~\cite{kivshar1998dark,kivsharbook2003,ledererPR2008discrete,kevrekidisBook2009} 
and Bose-Einstein condensates (BECs)~\cite{brazhnyiMPLB2004,morschRMP2006,dauxoisbook2006,BEC_book_2008,frantzeskakis2010dark,liu2018nonlinear}. 
A significant phenomenon associated with nonlinearity is the emergence of soliton solutions in the 
nonlinear Schr\"{o}dinger equation, arising from the balance 
between nonlinearity and dispersion~\cite{kivshar1998dark,kivsharbook2003,kevrekidisBook2009,dauxoisbook2006,BEC_book_2008}. 
Different types of nonlinearities give rise to different solitons.
For instance, focusing (attractive) nonlinearity typically leads to bright solitons,
while defocusing (repulsive) nonlinearity produces 
dark solitons. Unlike a bright soliton, a dark soliton corresponds to a defect with a sharp phase
change. Due to this unique characteristic, a dark soliton can host Majorana zero modes in topological fermionic superfluids~\cite{xu2014dark,liu2015soliton,zou2016traveling,zeng2019majorana}.
Both types of solitons have been 
experimentally observed in photonic systems~\cite{eisenbergPRL1998,morandotti2001self,fleischerNat2003,mandelik2004gap} and BECs~\cite{burger1999dark,khaykovich2002formation,strecker2002formation}.

One of the most significant consequences in a topological phase is the emergence of quantized 
transport that is robust against disorder, such as the quantum Hall effect~\cite{klitzing1980new,thouless1982quantized,simon1983holonomy}. This type of transport is characterized by
a topological invariant, such as the Chern number. 
The quantized transport in the quantum Hall effect
can be understood as the quantized pumping of electrons through a slow variation of a magnetic flux, 
a process known as Laughlin's pumping~\cite{laughlin1981quantized}. Alternatively, the Chern number also determines electron pumping
that occurs when a system parameter in a one-dimensional (1D) system is varied slowly over a period~\cite{thoulessPRB1983,niu1984quantised}.
Recently, the intersection of topology and nonlinearity has led to various intriguing phenomena, 
such as edge solitons~\cite{ablowitzPRA2014,leykamPRL2016,mukherjeePRX2021,tao2020hinge} and nonlinearity induced topological phase transition~\cite{maczewskySci2020,soneNP2024}.
In particular, it has been found that bright solitons exhibit quantized
pumping when a system parameter is adiabatically changed over a cycle~\cite{jurgensenNat2021,jurgensenPRL2022,fuPRL2022,fu2022twoD,mostaanNC2022,tuloupNJP2023,citro2023thouless,
	HuNJP2024,cao2024nonlinear,lyuPRR2024,szameit2024discrete,cao2025transport}. This phenomenon can 
be understood as the evolution of an instantaneous Wannier function with respect to the system parameter~\cite{jurgensenPRL2022,mostaanNC2022,fuPRL2022}.
Since the change in the center of mass of the Wannier function is determined by the Chern number,
the quantized nonlinear transport is inherently topological.

\begin{figure*}[tp]
	\includegraphics[width=1\linewidth]{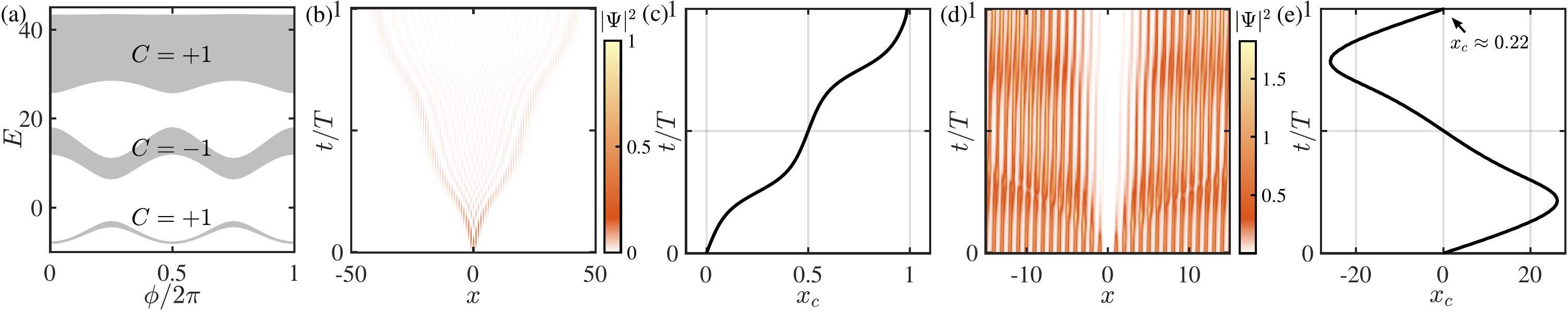}
	\caption{{Pumping of a hole on a continuous background and a wave packet without nonlinearity.} 
		{(a)} Linear Bloch band spectrum of the continuous model in Eq.~(\ref{dlGP}) with respect to a system parameter $\phi$. 
		The three lowest bands exhibit Chern numbers $C=\{1,-1,1\}$. 
		(b) and (d) The 
		squared amplitude of the wave function $\Psi(x,t)$ obtained by evolving an initial state based on the 
		Hamiltonian in Eq.~(\ref{dlGP}) by setting $g=0$.
		The initial state is chosen as a wave packet in (b) and a $\pi$-kink wave function in (d), resembling a bright 
		soliton and dark soliton, respectively (see the Supplemental Material for how to construct the initial state).		 
		Due to dispersion, the hole and the wave packet spread significantly during time evolution. 
		(c) and (e) The trajectory of the relative center-of-mass positions $x_c$ of the evolving state 
		in (b) and (d), respectively,
		defined by $x_c(t)=\int_{-L/2}^{L/2} x|\Psi(x,t)|^2 dx/\int_{-L/2}^{L/2}x|\Psi(x-1,0)|^2dx$, where $L$ is the system
		size. 
		We note that the center-of-mass position of an amplitude dip depends on the system size because the background 
		contribution dominates, making it ill-defined. To address this, we here consider the relative one compared to the case 
		where an initial hole is shifted by one unit cell. We see that the displacement per cycle is approximately $0.22$ for a hole, 
		whereas for a wave packet, it is very close to 1, matching the lowest-band Chern number.
		Here, we set $V_1=V_2=15$, $p_1=1/2$, $p_2=1$, and the period $T=100$.} 
	\label{fig1}
\end{figure*}

This raises a natural question: can a dark soliton also be transported in a quantized manner via 
Thouless pumping? The answer remains uncertain, as dark solitons significantly differ 
from bright ones. Unlike bright solitons, which can be perceived as Wannier functions,
dark solitons, characterized by a density dip, do not fit this description. Additionally, in a linear 
Thouless pumping system, an initial wave packet state of a Wannier function can undergo quantized 
transport for its center of mass, even as it becomes broader~\cite{jurgensenNat2021} [see Fig.~\ref{fig1}]. 
However, if the initial state is a hole on a continuous background with a 
sharp phase change, which resembles a 
dark soliton, then the center-of-mass position of the evolving state is not well defined [see Fig.~\ref{fig1}]. 
Finally, unlike bright solitons that occur with focusing nonlinearity and are typically stable, 
dark solitons arise with defocusing nonlinearity and appear in the energy gap above the first Bloch band~\cite{muryshev1999stability,fedichev1999dissipative,pelinovsky2008stability}, 
making them prone to instability and challenging to pump.

Despite the distinct behavior observed, we here predict that a dark soliton can 
exhibit topologically quantized transport when a system parameter is slowly varied 
over one period in a continuous system under optical lattices or within a tight-binding model. 
Our findings indicate that the center of the dark soliton closely follows the center of mass of an instantaneous Wannier function, 
suggesting that pumping is determined by the Chern number of the corresponding linear Bloch band. 
Additionally, we observe a fractional Thouless pumping of a dark soliton, namely, that a soliton is transported 
by half a unit cell over one period. In this scenario, the dark soliton corresponds to the flow of an 
instantaneous multi-band Wannier function. Our predictions can be experimentally confirmed in 
both ultracold atomic gases and photonic systems.

\emph{Continuous model}---To demonstrate Thouless pumping of dark solitons, we start by considering a 1D 
condensate in an optical lattice, whose dynamics is described by the dimensionless Gross-Pitaevskii (GP) equation,
\begin{equation}
	\label{dlGP}
	i\frac{\partial}{\partial t} \Psi(x,t)=\left[H_{\mathrm{L}}(x,\phi)+g|\Psi(x,t)|^2\right]\Psi(x,t).
\end{equation}
Here, $\Psi(x,t)$ is a wave function, and 
$H_{\mathrm{L}}(x,\phi)=-(1/2)\partial_x^2+V(x,\phi)$
is the linear Hamiltonian, where 
$V(x,\phi)$ is a $\phi$-dependent superlattice potential, taking the form~\cite{wang2013topological,nakajima2016topological,lohse2016thouless}
\begin{equation}
	\label{Vxt}
	V(x,\phi)=-V_1\cos\left(\frac{2\pi x}{p_1}\right)-V_2\cos\left(\frac{2\pi x}{p_2}-\phi\right)
\end{equation}
with $V_{1,2}$ and $p_{1,2}$ being dimensionless amplitudes and spatial periods of these two constitutive 
lattice potentials, respectively. The dynamical variation of the superlattice potential is controlled by the 
parameter $\phi$, which varies linearly with time as $\phi=2\pi t/T$ with $T$ being the modulation period.
The units of length, energy, and time are chosen as $a$, $\hbar^2/(ma^2)$, and $ma^2/\hbar$, respectively,
where $a$ is the spatial period of the potential $V(x,\phi)$ and $m$ is the atomic mass. 
The parameter $g$ represents the nonlinear coefficient. 
To ensure the stable existence of dark soliton solutions~\cite{muryshev1999stability,fedichev1999dissipative,pelinovsky2008stability}, we consider repulsive interactions 
with $g>0$ and set $g=1$ without loss of generality.

\begin{figure*}[t]
	\includegraphics[width=1\linewidth]{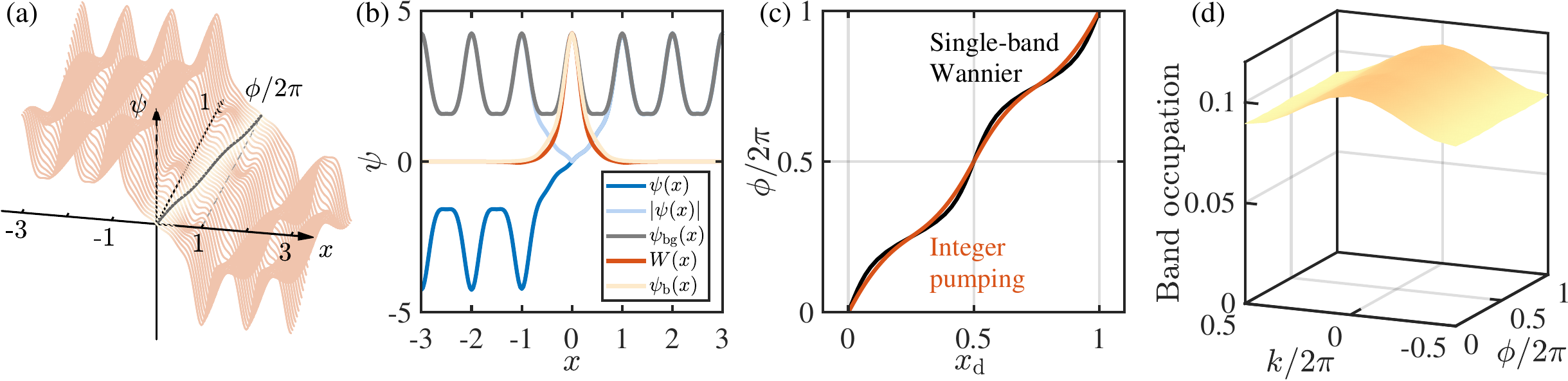}
	\caption{{Integer pumping of dark solitons and its underlying mechanism.} 
		(a) Evolution of an instantaneous dark soliton obtained by solving the stationary solution to 
		Eq.~(\ref{dlGP}) at a fixed $\phi$, illustrating integer transport of a dark soliton over one cycle.
		The gray line represents the trajectory of the center-of-mass positions of dark solitons, which
		are consistent with the positions of the defect where the phase changes abruptly.
		(b) The profile of the dark soliton wave function $\psi(x)$ (blue line), 
		its amplitude $|{\psi}(x)|$ (light blue line), the background wave function $\psi_{\mathrm{bg}}(x)$ 
		(gray line), the Wannier function of the lowest band $W(x)$ (red line), and the 
		mapped bright soliton wave function $\psi_{\mathrm{b}}(x)$ (light red line) at $\phi=0$. 
		We see that $\psi_{\mathrm{b}}(x)$ resembles $W(x)$ closely.  
		(c) The trajectories of the center-of-mass position of the dark soliton (red line) as well as 
		the lowest-band Wannier center (black line) with respect to $\phi$ over one period. 
		(d) Occupations of the mapped bright soliton $\psi_{\mathrm{b}}(x)$ 
		on the linear Bloch state at each momentum $k$ and $\phi$, illustrating nearly uniform
		occupation. In the basis of Wannier functions, $\psi_{\mathrm{b}}(x)$ 
		is predominantly composed of a single Wannier function.		 
		Here, we set $V_1=V_2=15$, $p_1=1/2$, $p_2=1$, and $\mu=4$ at $\phi=0$.}
	\label{fig2}
\end{figure*}

In the absence of nonlinearity, Thouless pumping for electrons that fill an entire Bloch band of the linear 
Hamiltonian $H_{\mathrm{L}}(\phi)$ is determined by
the Chern number of this band~\cite{thouless1982quantized,thoulessPRB1983,xiao2010berry},
\begin{align}
	\label{Chern}
	C_n=&\frac{1}{2\pi}\int_0^{2\pi}d\phi\int_0^{2\pi}dk \Omega_n(k,\phi),
\end{align}
where $\Omega_n(k,\phi)=i\left[\langle\partial_k u_{nk} |\partial_\phi u_{nk}\rangle - \langle\partial_\phi u_{nk} |\partial_k u_{nk}\rangle
 \right]$ is the Berry curvature with respect to
$k$ and $\phi$, which plays the role of another momentum.
$u_{nk}(x,\phi)$ is the periodic part of the Bloch state $\varphi_{nk}(x,\phi)$ at the Bloch momentum $k$ for the $n$th band.
By setting $V_1=V_2=15$, $p_1=1/2$, and $p_2=1$, we find that the three lowest Bloch bands exhibit the Chern number
of $1$, $-1$ and $1$, respectively, as illustrated in Fig.~\ref{fig1}(a).
Figure~\ref{fig1} also shows that without nonlinearity, while the center of mass of an initial wave packet can undergo quantized transport 
as $\phi$ is slowly varied over one period, an initial hole on a continuous background cannot achieve this transport.

In the presence of nonlinearity, since in the adiabatic limit with sufficiently large period $T$, the evolving state corresponds to 
an instantaneous eigenstate of the instantaneous nonlinear Hamiltonian~\cite{kivsharRMP1989,bandPRA2002,bandPRA2002_2,liuPRL2003,wuPRL2005},
we compute the instantaneous stable $\pi$-kink dark soliton solutions $\psi(x,\phi)$, also referred to as black solitons~\cite{kivshar1998dark}, 
at each $\phi$ via Newton's method. 
Here, $\psi(x,\phi)$ is a dark soliton 
solution to the nonlinear Hamiltonian at $\phi$ so that
$
	\mu\psi(x,\phi)=H_{\mathrm L}(\phi)\psi(x,\phi)+g|\psi(x,\phi)|^2\psi(x,\phi)
$
with $\mu$ being the nonlinear eigenenergy.
We find that the instantaneous solutions maintain 
spatial localization throughout the entire evolution, and return to its original shape after each cycle,
leading to transport by one unit cell, as illustrated by the trajectory of the defects across which the phase
suddenly changes in Fig.~\ref{fig2}(a). 
In addition, we follow the convention to define a squared amplitude of a wave packet 
$n_{\mathrm{d} }=|\psi_{\rm bg}(x,\phi)|^2-|\psi(x,\phi)|^2$ to characterize the hole of the dark soliton~\cite{parker2003deformation},
where $\psi_{\rm{bg}}(x,\phi)$ is the background wave function satisfying the nonlinear equation
$
	\mu\psi_{\rm{bg}}(x,\phi)=H_{\mathrm L}(\phi)\psi_{\rm{bg}}(x,\phi)+g|\psi_{\rm{bg}}(x,\phi)|^2\psi_{\rm{bg}}(x,\phi).
$
The background wave function [e.g., the gray line in Fig.~\ref{fig2}(b)] is actually the nonlinear Bloch state at $k=0$.
We subsequently define the center-of-mass position of the dark soliton by
$
\label{x_d}
x_{\mathrm{d}}(\phi)= \int x n_{\mathrm{d} } dx / \int n_{\mathrm{d} } dx
$, 
showing clearly that it exhibits a smooth forward shift by one unit cell per cycle [see Fig.~\ref{fig2}(c)].

In addition, in contrast to bright solitons that increasing attractive nonlinearity
always leads to zero pumping~\cite{jurgensenNat2021}, 
we find that stronger repulsive nonlinearity makes dark solitons unstable, thereby preventing the 
emergence of the trapped phenomenon.

\emph{Mechanism of quantized pumping}---Our results indicate that the dark soliton's transport aligns with 
the Chern number of the underlying linear Bloch band, similar to a bright soliton.
A bright soliton can be understood as a superposition of all Bloch states 
within a single band, making it analogous to a Wannier function~\cite{jurgensenPRL2022,mostaanNC2022,fuPRL2022}. This resemblance to 
a fully occupied band of electrons naturally leads to quantized Thouless pumping. 
A dark soliton, however, exhibits a phase kink that inherently breaks periodicity, preventing 
its construction from conventional Bloch states under periodic boundary conditions.

Another key distinction of dark solitons from bright solitons lies in their 
background wave function. This background corresponds to the nonlinear Bloch
state at $k=0$, which, in the weak nonlinearity regime, can be expressed as a 
superposition of Wannier functions from the corresponding linear Bloch states, i.e.,
$\psi_{\rm{bg}}\sim \sum_j W_{1j}(x)$, where $W_{1j}(x)$ represents the Wannier function 
of the lowest Bloch band at the $j$th unit cell. 
As the parameter $\phi$ is adiabatically varied over one cycle, these Wannier functions 
undergo transport with a displacement governed by
the Chern number of the relevant linear band. Consequently, the motion of the dark soliton  
embeded in the background can be interpreted as being driven by the background's dynamics.
While this framework successfully explains the observed integer Thouless pumping of dark solitons,
it inherently excludes the possibility of fractional pumping---a phenomenon we will
demonstrate in the following section. This limitation indicates that the current explanation 
remains incomplete and suggests the need for a more comprehensive theoretical understanding.

\begin{figure*}[htp]
	\includegraphics[width=1\linewidth]{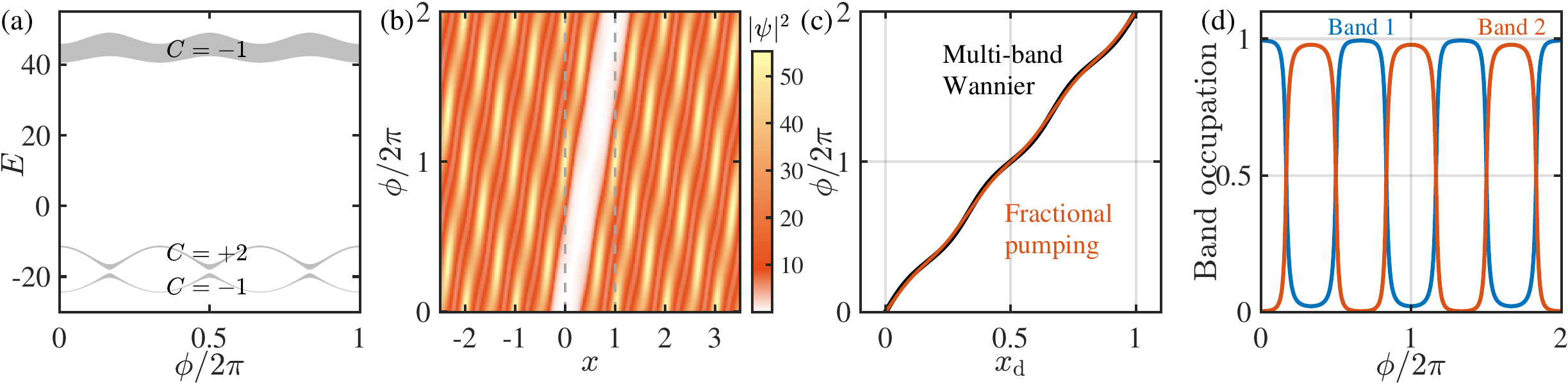}
	\caption{{Fractional Thouless pumping of dark solitons in the continuous model.} 
		{(a)} The linear band structure of the continuous Hamiltonian $H_{\mathrm{L}}$ 
		with two lowest bands close to each other. 
		{(b)} Evolution of the density distribution $|\psi|^2$ of instantaneous dark solitons 
		with respect to $\phi$, illustrating transport of the soliton by one unit cell over two periods. 
		The gray dashed lines indicate the initial and terminal positions of the dark soliton.
		{(c)} The trajectories of the center-of-mass positions of the dark soliton (red line)
		and the instantaneous multi-band Wannier functions of the two lowest bands (black line)
		with respect to $\phi$. 
		{(d)} Occupations ($\sum_k P_n(k,\phi)$ with $n=1,2$) of the mapped bright soliton $\psi_{\mathrm{b}}(x)$ 
		on the two lowest linear Bloch bands at each $\phi$, illustrating oscillating 
		band occupations. 
		Here, we set $V_1=20$, $V_2=60$, $p_1=1/3$, $p_2=1/2$, and $\mu=13$ at $\phi=0$.
	}
	\label{fig3}
\end{figure*}

In the following, we develop a framework to explain the observed phenomenon by mapping the 
dark soliton to a wave packet and demonstrating that it corresponds to a bright soliton solution for a 
modified nonlinear equation. 
The quantized transport of the dark soliton can thus be understood as
the transport of the bright soliton.  
Specifically, we map the dark soliton to a bright soliton via
\begin{equation} \label{Map-Eq}
	\psi_{\mathrm{b}}(x,\phi)=\psi_{\mathrm{bg}}(x,\phi)-|{\psi}(x,\phi)|.
\end{equation} 
We find that when $x$ is slightly away from the dark soliton's dip, $\psi_{\mathrm{b}}$ satisfies the 
following nonlinear equation (see Supplemental Material for derivation): 
\begin{equation}
	\label{NE_dip}
	\mu\psi_{\mathrm b}= \left(H_{\mathrm L}+3g \psi_{\mathrm{bg}}^2 \right)\psi_{\mathrm b}
	+g \left(\psi_{\mathrm b}^3 -3\psi_{\mathrm{bg}} \psi_{\mathrm b}^2  \right).
\end{equation} 
The $\psi_{\mathrm{b}}$ exhibits a clear similarity to the Wannier function
of the lowest band of the linear Hamiltonian $H_{\mathrm{L}}$ [see Fig.~\ref{fig2}(b)].
This similarity is further supported by the nearly uniform occupation [Fig.~\ref{fig2}(d)] of the bright soliton wave function 
at each $k$ and $\phi$ on the $n$th linear band ($n=1$) with the occupation defined as
$
	P_n(k,\phi)= |\langle\varphi_{nk}(\phi)|\psi_{\mathrm{b}}(\phi)\rangle|^2 /\langle\psi_{\mathrm{b}}(\phi)|\psi_{\mathrm{b}}(\phi)\rangle.
$
Thus, the motion of the bright soliton, and consequently the dark soliton, follows instantaneous linear 
Wannier functions during time evolution [see Fig.~\ref{fig2}(c)]. 
As the displacement of the linear Wannier center over one cycle is equal to the Chern number of the 
corresponding band, the quantized pumping of the dark soliton is dictated by the Chern number,
irrelevant to parameter details (see Supplemental Material).

\emph{Fractional pumping of dark solitons}---We now demonstrate that a dark soliton can undergo
quantized fractional transport over one cycle.
For the continuous model, we consider the optical lattices described in Eq.~(\ref{Vxt}) with parameters $V_1=20$, 
$V_2=60$, $p_1=1/3$, and $p_2=1/2$, resulting in three lowest linear bands with 
Chern numbers of $-1$, $2$, and $-1$, respectively [see Fig.~\ref{fig3}(a)]. 
The instantaneous dark solitons in the system exhibit transport of one unit cell over two cycles, or
half a unit cell over one cycle, as illustrated in Fig.~\ref{fig3}(b).
This fractional pumping is also clearly seen by the trajectory of the dark soliton's 
center-of-mass position [red line in Fig.~\ref{fig3}(c)]. 
We have previously noted that fractional transport cannot be interpreted as 
background dragging, as this would only result in integer pumping. 
To explain the phenomenon, we map the dark soliton to a bright soliton based on Eq.~(\ref{Map-Eq}).
We find that the bright soliton occupies the two lowest linear bands, with band occupations 
exhibiting alternating oscillations during evolution [see Fig.~\ref{fig3}(d)]. 
This indicates that the bright soliton and consequently the dark soliton follow the corresponding 
instantaneous linear multi-band Wannier function with fractional pumping of $(C_1+C_2)/2=1/2$ per cycle~\cite{jurgensenNP2023},
as shown in Fig.~\ref{fig3}(c).

\emph{Discussion and outlook}---We have predicted Thouless pumping of dark solitons, which can be experimentally
observed in both ultracold atomic gases and waveguide arrays. For the former, we consider a 
$^7\text{Li}$ BEC with the $s$-wave scattering length $a_s\approx1.43$ nm under optical lattices with the spatial 
period $a=532$ nm and the characteristic length of a transverse harmonic trap $a_\perp \approx1$ $\mu$m~\cite{khaykovich2002formation,strecker2002formation}. 
The profile of dark solitons is governed by the number of particles per unit cell far from the kink, 
which is given by $\mathcal{N}=\frac{a^2_\perp g}{4a_s a}\int^{-L/2+1}_{-L/2}|\psi(x)|^2dx \approx2335$~\cite{perez1998bose}.
Under our numerical simulations, we confirm that a period of $T=31.5$ ms is sufficient for the adiabatic limit. 
In addition, we have found that similar phenomena are able to occur in the off-diagonal Aubry-Andr\'{e}-Harper (AAH) model as detailed in 
the Supplemental Material. 
Thus, the dark soliton's pumping can also be realized in a 1D array of nearest-neighbor coupled 
waveguides with defocusing Kerr nonlinearity. Under the paraxial approximation, the propagation of light is 
described by a discrete nonlinear Schr{\"o}dinger equation (see Supplemental Material). Now, the wave function
$\Psi_x$ depicts the envelope of electrical fields at site $x$, and $t$ denotes the propagation distance along 
waveguides. To achieve $\phi$-dependent couplings in the off-diagonal AAH model, the inter-waveguide 
spacing is modulated as a function of $t$. Applying a typical parameter $J_0=0.15$ mm$^{-1}$~\cite{jurgensenNat2021,jurgensenNP2023}, we confirm 
that a propagation distance of 6.67 m per cycle satisfies the adiabatic condition.
The robustness against weak disorder is discussed in the Supplemental Material.

Recent studies on bright solitons have revealed a breakdown in the correspondence between linear band topology 
and Thouless pumping~\cite{tao2024nonlinearity}. Moreover, nonlinearity-induced integer and fractional pumping of bright solitons has 
been theoretically predicted~\cite{tao2024nonlinearity,tao2024nonlinearity_frac}. These findings raise the intriguing possibility of exploring analogous phenomena in 
dark solitons. Additionally, our work can be extended to two-dimensional (2D) systems, offering a promising 
avenue to investigate vortex~\cite{BEC_book_2008} or ring dark soliton~\cite{kivshar1998dark} transport through 2D nonlinear Thouless pumping. 
In addition, our results may inspire further research into the transport of dark solitons with 
Majorana zero modes in topological Fermi superfluids~\cite{xu2014dark}.

\begin{acknowledgments}
	This work is supported by the National Natural Science Foundation of China (Grant No. 12474265, 11974201,
	12374247, and 11974235)
	and Innovation Program for Quantum Science and Technology (Grant No. 2021ZD0301604).
	Y. Zhang is also supported by the Shanghai
	Municipal Science and Technology Major Project (Grant
	No. 2019SHZDZX01-ZX04).
	We also acknowledge the support by center of high performance computing, Tsinghua University.
\end{acknowledgments}

\begin{widetext}
%\maketitle
\setcounter{equation}{0} \setcounter{figure}{0} \setcounter{table}{0} %
\renewcommand{\theequation}{S\arabic{equation}}
\renewcommand{\thefigure}{S\arabic{figure}}
\renewcommand{\bibnumfmt}[1]{[S#1]}
\renewcommand{\citenumfont}[1]{S#1}
%\section{Supplemental Material for Quantized Thouless Pumping of Dark Solitons}
In the Supplemental Material, we will elaborate on how a $\pi$-kink initial wave function
is constructed in Section S-1,
present details on how to map a dark soliton to a bright soliton in Section S-2, 
show that although the trajectory of a soliton may be different for distinct system parameters,
their displacements over one cycle are equal in Section S-3,
present quantized Thouless pumping of dark solitons in the off-diagonal AAH model in Section S-4,
provide an example showing that when the linear bands become topologically trivial,
a dark soliton does not exhibit pumping in Section S-5,
and finally show the stability of Thouless pumping of dark solitons against weak disorder in Section S-6.

\section{S-1. Construction of an initial $\pi$-kink state}
In the main text, we show the time evolution of an initial $\pi$-kink state by a linear
Hamiltonian without nonlinearity. In this section, we will elaborate on how this initial 
state is constructed.

Let $W(x)$ (red in Fig.~\ref{fig-sm-kink-linear}) be the Wannier function of the lowest linear Bloch band
at $\phi=0$ and $\varphi_{10}(x)$ (gray line in Fig.~\ref{fig-sm-kink-linear}) 
be the lowest-band Bloch state at momentum $k=0$ and $\phi=0$. 
We define the absolute value of $\pi$-kink wave function by $n_{\mathrm{kink}}(x)=\varphi_{10}(x)-W(x)$ 
(light blue line in Fig.~\ref{fig-sm-kink-linear}).
To ensure that the value at the hole's center equals zero, we multiply a constant to both $W(x)$ and $\varphi_{10}(x)$ so that
the peak values of both $W(x)$ and $\varphi_{10}(x)$ equal one.
The initial $\pi$-kink state is then defined by
$\psi_{\mathrm{kink}}(x) = n_{\mathrm{kink}}(x)$ when $x>0$ and
$\psi_{\mathrm{kink}}(x) = -n_{\mathrm{kink}}(x)$ when $x<0$ (see the dark blue line in Fig.~\ref{fig-sm-kink-linear}).
This initial state is thus closely related to the Wannier function.

In addition, to ensure the numerical simulation of time evolution in a periodic-boundary system, 
we introduce an initial wave function consisting of two kinks that are significantly separated in space so that
they have negligible effects to each other.

\begin{figure}[htp]
	\includegraphics[width=0.4\linewidth]{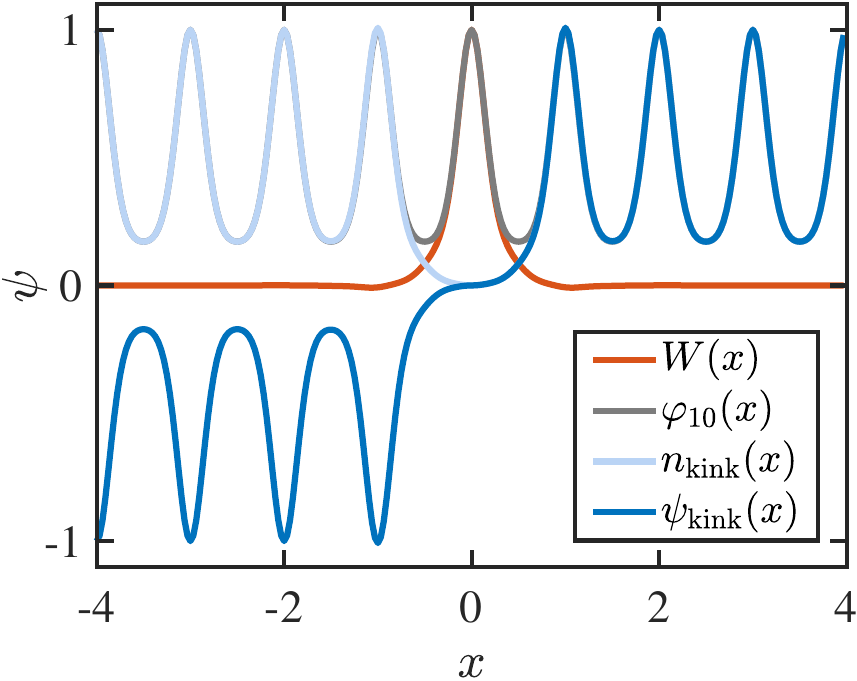}
	\caption{{The construction of an initial $\pi$-kink wave function at $\phi=0$.} 
		The light blue line represents the function of $n_{\mathrm{kink}}(x)=\varphi_{10}(x)-W(x)$, where 
		$W(x)$ is the Wannier function of the lowest linear band (red line) and
		$\varphi_{10}(x)$ is the lowest band Bloch state at momentum $k=0$ (gray line).
		The initial $\pi$-kink wave function $\psi_{\mathrm{kink}}(x)$ (dark blue line) is obtained
		by imposing a $\pi$ phase on the amplitude function  $n_{\mathrm{kink}}(x)$ when $x<0$.}
	\label{fig-sm-kink-linear}
\end{figure}

\section{S-2. Mapping a dark soliton to a wave packet}
To explain the observed Thouless pumping of a dark soliton, in this section, we will map a dark soliton to a wave packet through
\begin{equation}
	\psi_{\mathrm{b}}(x,\phi)=\psi_{\mathrm{bg}}(x,\phi)-\tilde{\psi}(x,\phi),
\end{equation} 
where $\tilde{\psi}(x,\phi) \equiv |{\psi}(x,\phi)|$, and derive a nonlinear equation that the function 
$\psi_{\mathrm{b}}(x,\phi)$ satisfies. 

We first write the time-dependent nonlinear GP equation in Eq.~(1) in the main text in the discrete form as 
\begin{equation}
	\label{GP_dis}
	i\frac{\partial}{\partial t}\Psi_n(t)=\sum_{n^\prime}\left[H_{\mathrm L}(\phi)\right]_{n,n^\prime}\Psi_{n^\prime}(t)+g|\Psi_n(t)|^2\Psi_n(t),
\end{equation}
which is obtained by dividing the continuous real space coordinate $x$ into grid points $x_n$ with 
$x_{n+1}-x_{n}=\Delta x$ ($n=1,2,\dots,N$). 
Here, $\Psi_n=\Psi(x_{n})$ and
\begin{align}
	\label{HL_mat}
	\left[H_{\mathrm L}(\phi)\right]_{n,n^\prime}=-\frac{1}{2\Delta x^2}\left(\delta_{n,n^\prime+1}+\delta_{n,n^\prime-1}\right)+\left[\frac{1}{\Delta x^2}+V(x_n,\phi)\right]\delta_{n,n^\prime},
\end{align}
where $V(x_n,\phi)$ is the superlattice potential at the grid point $x_n$. Writing $\Psi_n=\psi_n e^{-i\mu t}$,
we can obtain the stationary state $\psi_{n}$ by solving 
\begin{equation}
	\label{NE_dark}
	\mu\psi_n(\phi)=\sum_{n^\prime}\left[H_{\mathrm L}(\phi)\right]_{n,n^\prime}\psi_{n^\prime}(\phi)+g|\psi_n(\phi)|^2\psi_n(\phi),
\end{equation}
where $\mu$ is the nonlinear eigenenergy. For a $\pi$-kink solution with a sudden phase change, we derive 
the equation satisfied by its amplitude function $\tilde{\psi}_n$ (assume that the wave function takes negative
values on the left-hand side).
When $\psi_{n-1}>0$ or $\psi_{n+1}<0$, $\tilde{\psi}_n$ 
clearly also satisfies Eq.~(\ref{NE_dark}). 
However, when $\psi_{n+1}>0$ and $\psi_{n-1}<0$, we obtain
\begin{equation}
	\frac{{\psi}_{n+1} +{\psi}_{n-1} }{-2\Delta x^2}
	= \frac{\tilde{\psi}_{n+1} +\tilde{\psi}_{n-1} }{-2\Delta x^2} +\frac{\tilde{\psi}_{n-1}}{\Delta x^2}.
\end{equation}	
Based on this result, we derive that 
$\tilde{\psi}_n$ satisfies the following nonlinear equation, 
\begin{equation}
	\label{NE_modi}
	\mu\tilde{\psi}_n(\phi)=\sum_{n^\prime}\left[H_{\mathrm L}(\phi)+\Delta H\right]_{n,n^\prime}\tilde{\psi}_{n^\prime}(\phi)+g|\tilde{\psi}_n(\phi)|^2\tilde{\psi}_n(\phi),
\end{equation}
where an additional modified term $\Delta H$ is introduced. 
This term is nonzero only at positions 
where 
$\psi_{n+1}\ge0$ and $\psi_{n}<0$, and the discrete form is given by 
\begin{equation}
	\label{dH}
	[\Delta H]_{n,n+1}=[\Delta H]_{n+1,n}^*=-2[H_{\mathrm L}(\phi)]_{n,n+1}=\frac{1}{\Delta x^2}.
\end{equation}
We note that $\tilde{\psi}$ satisfies the periodic boundary conditions, i.e.,
$\tilde{\psi}_1=\tilde{\psi}_{N+1}$.  

The background wave function $\psi_{\rm bg}$ is a nonlinear Bloch state at $k=0$ 
of Eq.~(\ref{NE_dark}) that satisfies $\psi_{{\rm bg},n}=\psi_{{\rm bg},n+1/\Delta x}$. 
Based on Eq.~(\ref{NE_modi}) and Eq.~(\ref{NE_dark}) satisfied by $\psi_{\rm bg}$ and
the fact that $\tilde{\psi}_n=\psi_{{\rm bg},n}-\psi_{{\rm b},n}$, we derive that
\begin{equation}
	\label{NE_dip}
	\mu \psi_{{\rm b}, n}=\sum_{n^\prime}\left[H_{\mathrm L}(\phi)\right]_{n,n^\prime} \psi_{{\rm b}, n^\prime} 
	-\sum_{n^\prime}\left[\Delta H\right]_{n,n^\prime}(\psi_{{\rm bg},n^\prime}-\psi_{{\rm b},n^\prime})
	-3g\psi_{{\rm b}, n}^2 \psi_{{\rm bg},n}+g\psi_{{\rm b}, n}^3 +
	3g\psi_{{\rm bg},n}^2 \psi_{{\rm b}, n},
\end{equation}
where we have assumed that the wave function is real. We see that the wave packet $\psi_{{\rm b},n}$
is a solution to the above nonlinear equation. Since $\Delta H$ only takes effects near the kink, we consider
the following nonlinear equation without this term,
\begin{equation}
	\label{NE_dip_tidle}
	\tilde{\mu} \tilde{\psi}_{{\rm b}, n}=\sum_{n^\prime}\left[H_{\mathrm L}(\phi)\right]_{n,n^\prime} \tilde{\psi}_{{\rm b}, n^\prime} 
	-3g\psi_{{\rm b}, n}^2 \tilde{\psi}_{{\rm bg},n}+g\tilde{\psi}_{{\rm b}, n}^3 +
	3g\psi_{{\rm bg},n}^2 \tilde{\psi}_{{\rm b}, n}.
\end{equation}
For each $\phi$, we set $\psi_{\rm b}(\phi)$ as an initial input wave function and utilize Newton's method 
to solve the nonlinear eigenstate $\tilde{\psi}_{{\rm b}}(\phi)$ of Eq.~(\ref{NE_dip_tidle}) with the constraint of 
$\sum_n|\tilde{\psi}_{{\rm b},n}(\phi)|^2 \Delta x=\sum_n|\psi_{{\rm b},n}(\phi)|^2 \Delta x$. 

Figures~\ref{fig_sm_modi}(a) and (b) show that both $\psi_{\rm b}$ and $\tilde{\psi}_{{\rm b}}$
exhibit the Thouless pumping by one unit cell over one cycle, and during time evolution, their shape remains
similar.
For example, at $\phi=0$ and $\phi=\pi/2$, the two wave functions only differ near the kink, as 
illustrated in Figs.~\ref{fig_sm_modi}(c) and (d). 
We also plot the occupations of $\psi_{\rm b}(\phi)$ and $\tilde{\psi}_{{\rm b}}(\phi)$ on the linear bands 
as a function of $\phi$ in Figs.~\ref{fig_sm_modi}(e) and (f), showing that the lowest band dominates
the population for both the two cases.

\begin{figure}[htp]
	\includegraphics[width=0.7\linewidth]{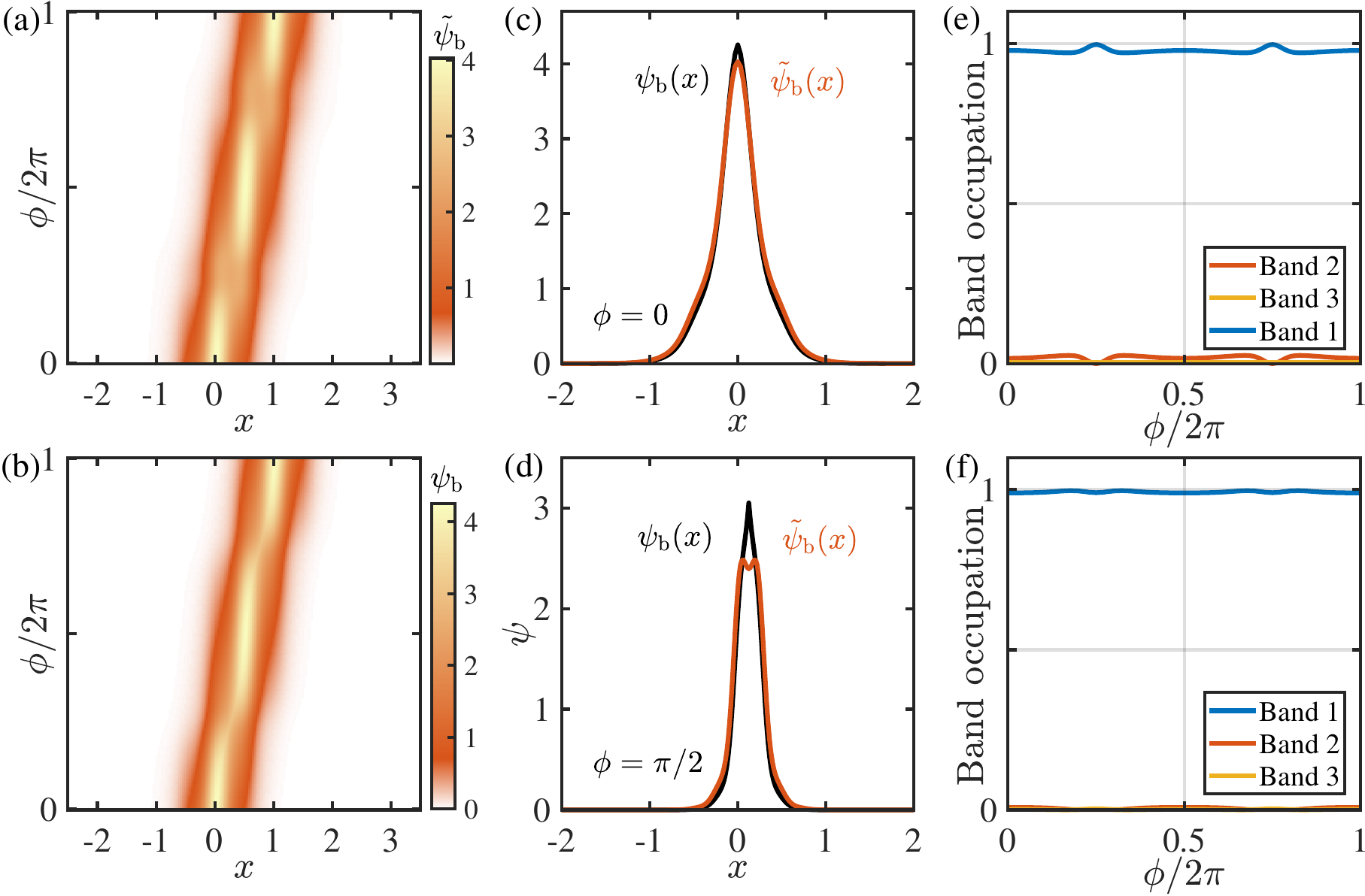}
	\caption{Comparison between the instantaneous nonlinear eigenstates $\tilde{\psi}_{\mathrm b}$ and $\psi_{\mathrm b}$. 
		{(a) and (b)} The evolution of $\tilde{\psi}_{\mathrm b}$ and  
		$\psi_{\mathrm b}$ with respect to $\phi$ over one cycle, respectively. 
		{(c) and (d)} The profiles of $\tilde{\psi}_{\mathrm b}$ (red line) and $\psi_{\mathrm b}$ (black line). 
		In (c), $\phi=0$, 
		and in (d), $\phi=\pi/2$. 
		(e) and (f) Occupations of $\tilde{\psi}_{\mathrm b}$ and $\psi_{\mathrm b}$ 
		on the three lowest linear bands at each $\phi$. 
		Here, we consider the superlattice potential $V(x,\phi)$ in Eq.~(1) in the main text with $V_1=V_2=15$, $p_1=1/2$, $p_2=1$, and $g=1$.}
	\label{fig_sm_modi}
\end{figure}

\section{S-3. Topological feature of pumping of dark solitons}
In this section, we will show that although the trajectory of a soliton may be different for distinct system parameters,
their displacements over one cycle are equal since the displacement is determined by the
Chern number of the corresponding linear bands.
Specifically, we consider the two cases with $\{V_1,V_2\}=\{15,15\}$ and $\{V_1,V_2\}=\{30,20\}$, 
whose lowest linear bands exhibit the Chern number of $1$. Figure~\ref{fig-sm-V1V2-change} illustrates that
although the trajectories of the dark solitons do not coincide, their displacements per cycle are identical, meaning
that one does not need to fine-tune system parameters to achieve quantized pumping.

\begin{figure}[htp]
	\includegraphics[width=0.3\linewidth]{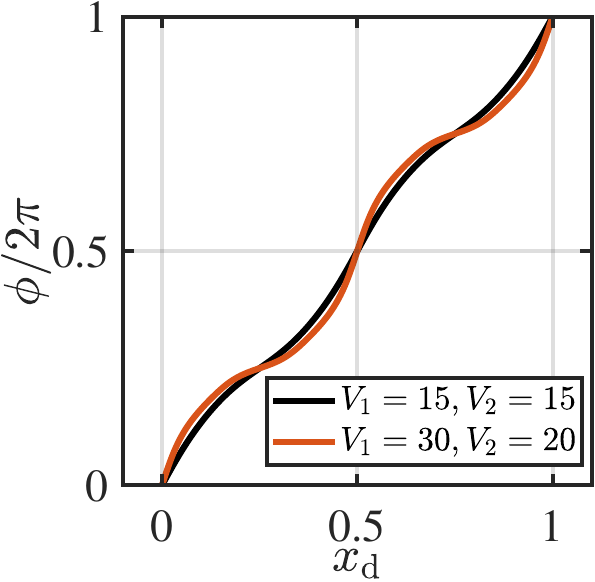}
	\caption{{Quantized pumping for different parameters.} The trajectories of the 
		center-of-mass position of the dark soliton for the case with $\{V_1,V_2\}=\{15,15\}$ (black line) and 
		$\{V_1,V_2\}=\{30,20\}$ (red line) over one cycle. Here, we set $p_1=1/2$, $p_2=1$. 
		For the black line, $\mu=4$ at $\phi=0$, and for the red line, $\mu=-2$ at $\phi=0$.}
	\label{fig-sm-V1V2-change}
\end{figure}

\begin{figure}[htp]
	\includegraphics[width=1\linewidth]{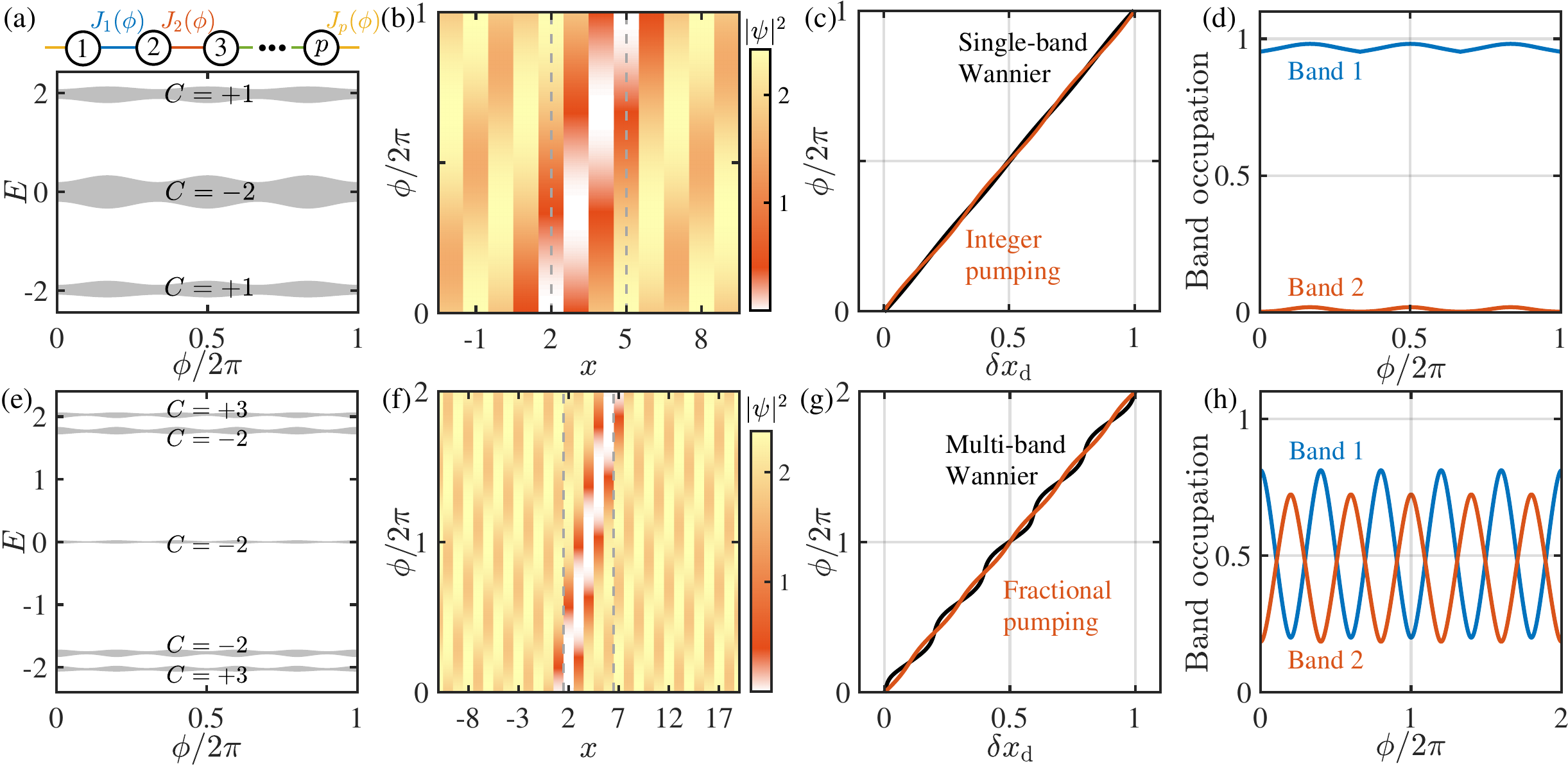}
	\caption{ {Integer (top row) and fractional (bottom row) Thouless pumping of dark solitons in the off-diagonal AAH model.} 
		(a) and (e) The linear Bloch band structure as a function of $\phi$. 
		We also presents schematics of the off-diagonal AAH model in (a).
		In each unit cell, there are $p$ sites with $\phi$-dependent 
		nearest-neighbor couplings $J_i(\phi)$ ($i=1,\dots,p$). The system is periodically modulated 
		in $\phi$ with period $2\pi$. 
		When $p=3$, $q=1$, $J_a=0.8$, and $\delta \phi=\pi/3$,
		the system exhibits three separated linear Bloch bands with Chern numbers $C=\{1,-2,1\}$ as shown in (a), 
		and when $\{p,q\}=\{5,2\}$, $J_a=0.95$, $\delta \phi=0$,
		there are five separated linear bands with Chern numbers $C=\{3,-2,-2,-2,3\}$ as shown in (e). 
		(b) and (f) Evolution of the density distribution $|\psi|^2$ of the discrete instantaneous dark solitons 
		with respect to $\phi$, illustrating that the soliton is pumped by one unit cell over (b) one cycle
		and (f) two cycles.
		The gray dashed lines denote the initial and terminal positions of the dark soliton. 
		(c) and (g) The trajectories of the relative center-of-mass position of the dark soliton (red line) and 
		the instantaneous Wannier function (black line) with respect to $\phi$.
		In (c) and (g), the lowest band Wannier function and the multi-band Wannier function  
		of the two lowest linear bands are considered, respectively. 
		Here $\delta x_{\mathrm d}$ is defined as 
		$\delta x_{\mathrm d}(\phi)=\left[x_{\mathrm d}(\phi)-x_{\mathrm d}(0)\right]/p$.
		(d) and (h) Occupations of the mapped discrete bright soliton on the two lowest linear bands with respect to $\phi$.
		(d) shows predominant occupation of the lowest band over one cycle and (h) shows alternating oscillations 
		over two cycles.
		In {(a)-(d)}, we set $p=3$, $q=1$, $J_a=0.8$, $\delta \phi=\pi/3$, and $\mu=0$ at $\phi=0$,
		and in (e)-(h), we set $\{p,q\}=\{5,2\}$, $J_a=0.95$, $\delta \phi=0$, and $\mu=0.1$ at $\phi=0$.}
	\label{fig-sm-TB}
\end{figure}

\section{S-4. Quantized Thouless pumping of dark solitons in a tight-binding model}
Solitons also widely exist in photonic systems, such as coupled waveguide arrays~\cite{SM_eisenbergPRL1998,SM_morandotti2001self,SM_fleischerNat2003,SM_mandelik2004gap,SM_keLPR2016}, which
are described by tight-binding models. To determine whether pumping of dark solitons can be observed in these systems, 
in this section, we consider the following dimensionless discrete nonlinear Schr{\"o}dinger equation:
\begin{equation}
	\label{NLSETB}
	i\frac{\partial}{\partial t} \Psi_x(t)=\sum_{x^\prime}H^{\textrm{lin}}_{x,x^\prime}(\phi)\Psi_{x^\prime}(t)+g|\Psi_x(t)|^2\Psi_x(t),
\end{equation}
where $\Psi_x(t)$ denotes the value of a wave function at site $x$ and time $t$, and $H^{\textrm{lin}}(\phi)$ is a $\phi$-dependent linear 
tight-binding Hamiltonian with a period of $2\pi$. Same as the continuous model, $\phi$ is linearly modulated 
in time as $\phi=2\pi t/T$ with sufficiently long period $T$.

To demonstrate the phenomena, we consider a 1D off-diagonal AAH model 
with $p$ sites per unit cell and nearest-neighbor hoppings~\cite{SM_AAH_1,SM_AAH_2,SM_krausPRL2012,SM_lang2012edge}, 
as illustrated in Fig.~\ref{fig-sm-TB}(a).
The hopping strength $J_{x}$ between sites $x$ and $x+1$ satisfies the condition 
$J_{x}=J_{[(x-1) \text{ mod } p]+1}$. 
We vary $J_{x}$ based on the equation 
$
J_{x}(\phi)=J_0+J_a \cos \left( \phi+\delta \phi-{2\pi q}(x-1)/{p}\right),
$
where $J_a$ is the amplitude of the oscillations, $J_0$ is the central hopping strength, 
and $\delta \phi$ is a phase shift term. The positive integer $q$ ($q<p/2$), coprime with $p$, 
depicts the relation between adjacent nearest-neighbor hoppings. 
For simplicity, we set $J_0=1$ as the unit of energy. 
Figure~\ref{fig-sm-TB}(a) also displays the band structure of the linear Hamiltonian $H^{\textrm{lin}}$, 
showing three bands with Chern numbers $C=\{1,-2,1\}$.

With the introduction of nonlinearity ($g=1$), we calculate instantaneous dark solitons as $\phi$ 
evolves from 0 to $2\pi$, demonstrating that the dark soliton is transported by one unit cell (three sites) 
over a single cycle, as illustrated in Fig.~\ref{fig-sm-TB}(b), 
similar to the continuous case.
The transport of the dark soliton is more clearly observed through their center-of-mass positions,
which follow the Wannier functions of the lowest linear band during 
evolution, as shown in Fig.~\ref{fig-sm-TB}(c).
Similarly, we map the dark soliton to a bright soliton, finding that the bright soliton closely resembles
the Wannier functions as revealed by nearly uniform band occupations 
of the mapped bright soliton on the linear bands at each $\phi$ in Fig.~\ref{fig-sm-TB}(d). 
This resemblance indicates that the displacement of the dark soliton remains 
governed by the Chern number of the corresponding Bloch band.

We now present fractional Thouless pumping of dark solitons in the off-diagonal AAH model. 
We consider the parameters $\{p,q\}=\{5,2\}$, $J_a=0.95$, and $\delta \phi=0$ so that five bands 
with Chern numbers $C=\{3,-2,-2,-2,3\}$ arise [see Fig.~\ref{fig-sm-TB}(e)]. 
Fig.~\ref{fig-sm-TB}(f) shows that an instantaneous discrete dark soliton is transported by one unit cell 
(five sites) over two cycles, corresponding to $1/2$ pumping per cycle. 
The fractional transport is more clearly illustrated by the trajectory of the relative center-of-mass position $\delta x_{\mathrm{d}}$
of the dark soliton [red line in Fig.~\ref{fig-sm-TB}(g)]. 
To explain the phenomenon, we map the discrete dark soliton to a bright one based on 
Eq.~(4) in the main text. Similar to the continuous case, the bright soliton bifurcates from the 
two lowest linear bands and its occupations show alternating oscillations during evolution [see Fig.~\ref{fig-sm-TB}(h)]. 
This implies that the bright soliton, and consequently the discrete dark soliton follow the instantaneous linear multi-band 
Wannier function with $(C_1+C_2)/2=1/2$ pumping per cycle [black line in Fig.~\ref{fig-sm-TB}(g)].

\begin{figure}[htp]
	\includegraphics[width=1\linewidth]{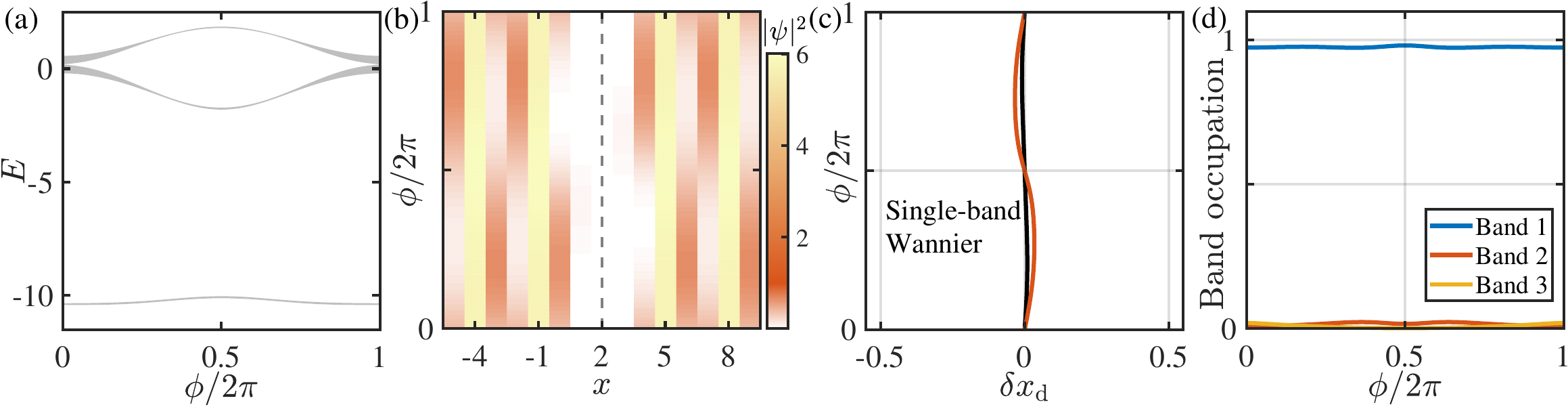}
	\caption{{Absence of pumping of dark solitons under trivial linear topology.} 
		{(a)} Linear Bloch band spectrum of the off-diagonal AAH model as a function of $\phi$, 
		illustrating three bands with zero Chern numbers. 
		{(b)} Evolution of the density distribution $|\psi|^2$ of instantaneous dark solitons over one cycle. 
		The two gray dashed lines representing the initial and terminal positions of the dark soliton
		coincide with each other. 
		{(c)} The trajectories of the relative center-of-mass position of the dark soliton $\delta x_{\mathrm{d}}$ (red line) and the 
		lowest band Wannier function (black line) with respect to $\phi$. {(d)} Occupations of the mapped bright soliton 
		on the three lowest linear bands at each $\phi$. 
		Here, we set $p=3$, $q=1$, $J_a=0.8$, $\delta \phi=\pi/3$, $\{m_x\}=\{0,-10,0\}$ with $x=1,2,3$, and $\mu=-5$ at $\phi=0$.}
	\label{fig-sm-TB-trivial}
\end{figure}

\section{S-5. Absence of pumping of dark solitons for topologically trivial linear bands}
In this section, we will provide an example showing that when the linear bands become topologically trivial,
a dark soliton does not exhibit a pumping over one period. 

We consider the off-diagonal AAH model and introduce an additional on-site potential $m_x$ with a 
period of $p$ sites to trivialize the topology. By setting $\{m_x\}=\{0,-10,0\}$ with $x=1,2,3$, 
we obtain three bands with zero Chern numbers [see Fig.~\ref{fig-sm-TB-trivial}(a)]. 
We find that the instantaneous dark soliton returns to the initial position over one cycle, 
as illustrated in Fig.~\ref{fig-sm-TB-trivial}(b). The zero displacement is also confirmed by the trajectory 
of the relative center-of-mass position of the dark soliton $\delta x_{\mathrm{d}}$ during evolution 
[red line in Fig.~\ref{fig-sm-TB-trivial}(c)]. To verify its topological mechanism, we calculate band occupations 
of the mapped bright soliton on the linear bands at each $\phi$ and find that 
this bright soliton mainly bifurcates from the lowest band as shown in Fig.~\ref{fig-sm-TB-trivial}(d), 
indicating that the bright soliton and consequently the dark soliton follow the instantaneous lowest-band 
Wannier function over a single cycle [black line in Fig.~\ref{fig-sm-TB-trivial}(c)].

\section{S-6. Robustness against weak disorder}
In this section, we will introduce disorder during time evolution in both continuous and tight-binding models, and 
demonstrate the robustness of Thouless pumping of dark solitons against weak disorder. 
Specifically, we introduce a random potential at each time and spatial position, 
expressed as $V_{\rm d}(x,t)=W_{\rm d}U_{\rm d}(x,t)$, where $W_{\rm d}$ denotes the disorder strength 
and $U_{\rm d}(x,t)$ represents an uncorrelated uniform random distribution within $\left[-1,1\right]$.

Without loss of generality, we consider the integer case. Figure~\ref{fig-sm-disorder}(b) illustrates 
that the dynamics of a dark soliton in the presence of disorder does not exhibit a clear difference from
the disorder-free case shown in Fig.~\ref{fig-sm-disorder}(a).
The stability is more clearly demonstrated by comparing the initial and final density 
distributions in Fig.~\ref{fig-sm-disorder}(c). 
Despite the presence of disorder, the dark soliton is transported by one unit cell over one cycle 
and maintains a profile nearly identical to the input wave function.
Similar conclusions are drawn for the tight-binding model, as depicted in Figs.~\ref{fig-sm-disorder}(d)-(f).

\begin{figure}[htp]
	\includegraphics[width=0.8\linewidth]{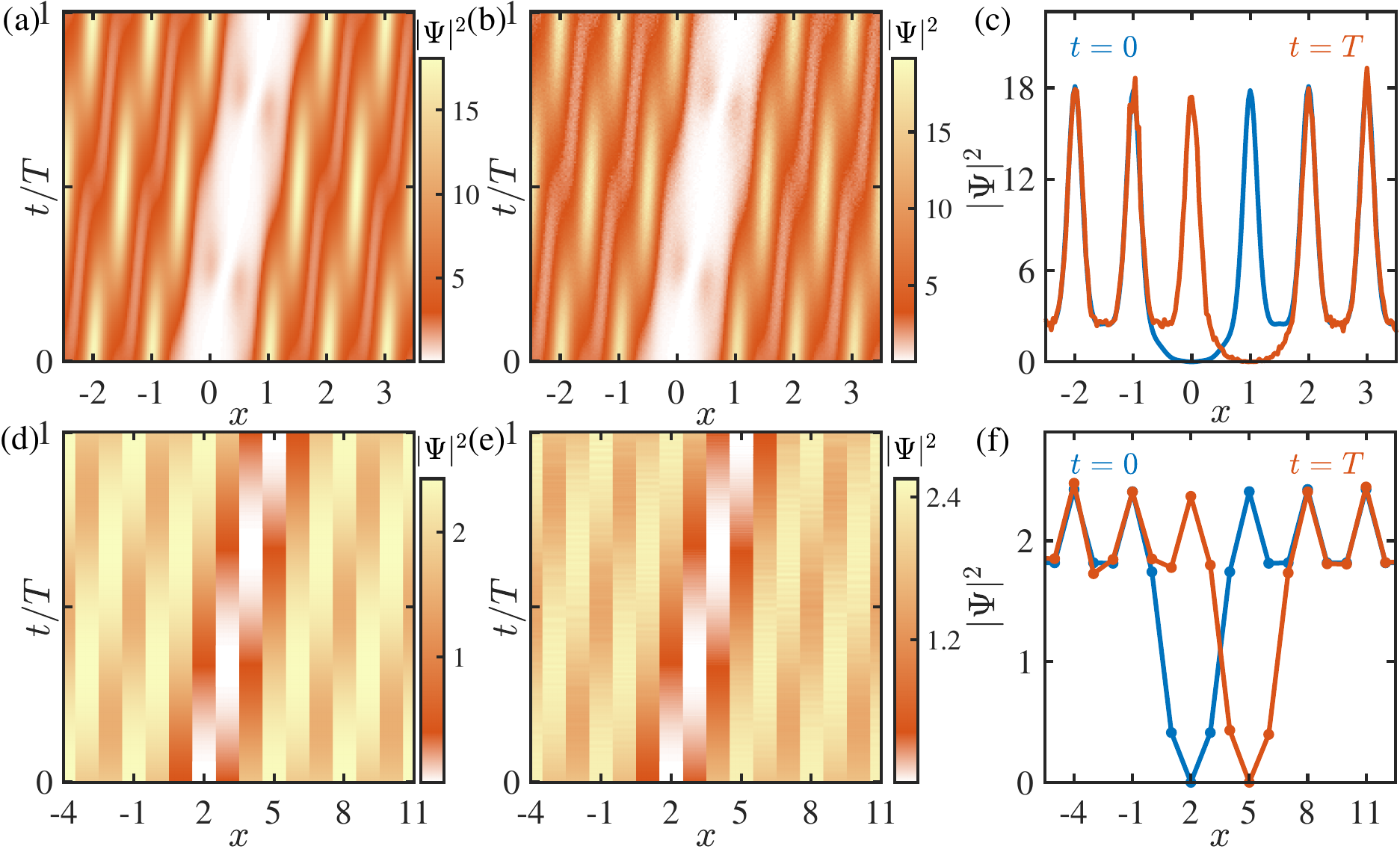}
	\caption{{Time evolution of dark solitons with disorder in continuous and tight-binding models.} 
		{(a)} Time evolution of the density distribution $|\Psi|^2$over one cycle in the continuous model without disorder. 
		{(b)} The same as \textbf{a} but with the disorder strength $W_{\rm d}=0.1$. 
		{(c)} Comparison between the profiles of $|\Psi|^2$ at time $t=0$ (blue line) and $t=T$ (red line), where $T$
		is the evolution time. 
		{(d)-(f)} The same as (a)-(c) but in the tight-binding model. In (e) and (f), 
		$W_{\mathrm d}=0.01$. Here, for the continuous model in (a)-(c), 
		we set $V_1=V_2=15$, $p_1=1/2$, $p_2=1$, $\mu=4$ at $\phi=0$, and $T=1000$, and
		for the tight-binding model in (d)-(f), we set $p=3$, $q=1$, $J_a=0.8$, $\delta \phi=\pi/3$, $\mu=0$ at $\phi=0$, and $T=1000$.}
	\label{fig-sm-disorder}
\end{figure}

\end{widetext}


\begin{thebibliography}{62}%
	\makeatletter
	\providecommand \@ifxundefined [1]{%
		\@ifx{#1\undefined}
	}%
	\providecommand \@ifnum [1]{%
		\ifnum #1\expandafter \@firstoftwo
		\else \expandafter \@secondoftwo
		\fi
	}%
	\providecommand \@ifx [1]{%
		\ifx #1\expandafter \@firstoftwo
		\else \expandafter \@secondoftwo
		\fi
	}%
	\providecommand \natexlab [1]{#1}%
	\providecommand \enquote  [1]{``#1''}%
	\providecommand \bibnamefont  [1]{#1}%
	\providecommand \bibfnamefont [1]{#1}%
	\providecommand \citenamefont [1]{#1}%
	\providecommand \href@noop [0]{\@secondoftwo}%
	\providecommand \href [0]{\begingroup \@sanitize@url \@href}%
	\providecommand \@href[1]{\@@startlink{#1}\@@href}%
	\providecommand \@@href[1]{\endgroup#1\@@endlink}%
	\providecommand \@sanitize@url [0]{\catcode `\\12\catcode `\$12\catcode
		`\&12\catcode `\#12\catcode `\^12\catcode `\_12\catcode `\%12\relax}%
	\providecommand \@@startlink[1]{}%
	\providecommand \@@endlink[0]{}%
	\providecommand \url  [0]{\begingroup\@sanitize@url \@url }%
	\providecommand \@url [1]{\endgroup\@href {#1}{\urlprefix }}%
	\providecommand \urlprefix  [0]{URL }%
	\providecommand \Eprint [0]{\href }%
	\providecommand \doibase [0]{https://doi.org/}%
	\providecommand \selectlanguage [0]{\@gobble}%
	\providecommand \bibinfo  [0]{\@secondoftwo}%
	\providecommand \bibfield  [0]{\@secondoftwo}%
	\providecommand \translation [1]{[#1]}%
	\providecommand \BibitemOpen [0]{}%
	\providecommand \bibitemStop [0]{}%
	\providecommand \bibitemNoStop [0]{.\EOS\space}%
	\providecommand \EOS [0]{\spacefactor3000\relax}%
	\providecommand \BibitemShut  [1]{\csname bibitem#1\endcsname}%
	\let\auto@bib@innerbib\@empty
	%</preamble>
	\bibitem [{\citenamefont {Kivshar}\ and\ \citenamefont
		{Luther-Davies}(1998)}]{kivshar1998dark}%
	\BibitemOpen
	\bibfield  {author} {\bibinfo {author} {\bibfnamefont {Y.~S.}\ \bibnamefont
			{Kivshar}}\ and\ \bibinfo {author} {\bibfnamefont {B.}~\bibnamefont
			{Luther-Davies}},\ }\bibfield  {title} {\bibinfo {title} {Dark optical
			solitons: physics and applications},\ }\href
	{https://doi.org/10.1016/S0370-1573(97)00073-2} {\bibfield  {journal}
		{\bibinfo  {journal} {Phys. Rep.}\ }\textbf {\bibinfo {volume} {298}},\
		\bibinfo {pages} {81} (\bibinfo {year} {1998})}\BibitemShut {NoStop}%
	\bibitem [{\citenamefont {Kivshar}\ and\ \citenamefont
		{Agrawal}(2003)}]{kivsharbook2003}%
	\BibitemOpen
	\bibfield  {author} {\bibinfo {author} {\bibfnamefont {Y.~S.}\ \bibnamefont
			{Kivshar}}\ and\ \bibinfo {author} {\bibfnamefont {G.~P.}\ \bibnamefont
			{Agrawal}},\ }\href@noop {} {\emph {\bibinfo {title} {Optical Solitons: From
				Fibers to Photonic Crystals}}}\ (\bibinfo  {publisher} {Elsevier Science},\
	\bibinfo {year} {2003})\BibitemShut {NoStop}%
	\bibitem [{\citenamefont {Lederer}\ \emph {et~al.}(2008)\citenamefont
		{Lederer}, \citenamefont {Stegeman}, \citenamefont {Christodoulides},
		\citenamefont {Assanto}, \citenamefont {Segev},\ and\ \citenamefont
		{Silberberg}}]{ledererPR2008discrete}%
	\BibitemOpen
	\bibfield  {author} {\bibinfo {author} {\bibfnamefont {F.}~\bibnamefont
			{Lederer}}, \bibinfo {author} {\bibfnamefont {G.~I.}\ \bibnamefont
			{Stegeman}}, \bibinfo {author} {\bibfnamefont {D.~N.}\ \bibnamefont
			{Christodoulides}}, \bibinfo {author} {\bibfnamefont {G.}~\bibnamefont
			{Assanto}}, \bibinfo {author} {\bibfnamefont {M.}~\bibnamefont {Segev}},\
		and\ \bibinfo {author} {\bibfnamefont {Y.}~\bibnamefont {Silberberg}},\
	}\bibfield  {title} {\bibinfo {title} {Discrete solitons in optics},\ }\href
	{https://doi.org/10.1016/j.physrep.2008.04.004} {\bibfield  {journal}
		{\bibinfo  {journal} {Phys. Rep.}\ }\textbf {\bibinfo {volume} {463}},\
		\bibinfo {pages} {1} (\bibinfo {year} {2008})}\BibitemShut {NoStop}%
	\bibitem [{\citenamefont {Kevrekidis}(2009)}]{kevrekidisBook2009}%
	\BibitemOpen
	\bibfield  {author} {\bibinfo {author} {\bibfnamefont {P.~G.}\ \bibnamefont
			{Kevrekidis}},\ }\href@noop {} {\emph {\bibinfo {title} {The discrete
				nonlinear Schr{\"o}dinger equation: mathematical analysis, numerical
				computations and physical perspectives}}},\ Vol.\ \bibinfo {volume} {232}\
	(\bibinfo  {publisher} {Springer},\ \bibinfo {address} {Berlin, Heidelberg},\
	\bibinfo {year} {2009})\BibitemShut {NoStop}%
	\bibitem [{\citenamefont {Brazhnyi}\ and\ \citenamefont
		{Konotop}(2004)}]{brazhnyiMPLB2004}%
	\BibitemOpen
	\bibfield  {author} {\bibinfo {author} {\bibfnamefont {V.~A.}\ \bibnamefont
			{Brazhnyi}}\ and\ \bibinfo {author} {\bibfnamefont {V.~V.}\ \bibnamefont
			{Konotop}},\ }\bibfield  {title} {\bibinfo {title} {Theory of nonlinear
			matter waves in optical lattices},\ }\href
	{https://doi.org/10.1142/S0217984904007190} {\bibfield  {journal} {\bibinfo
			{journal} {Mod. Phys. Lett. B}\ }\textbf {\bibinfo {volume} {18}},\ \bibinfo
		{pages} {627} (\bibinfo {year} {2004})}\BibitemShut {NoStop}%
	\bibitem [{\citenamefont {Morsch}\ and\ \citenamefont
		{Oberthaler}(2006)}]{morschRMP2006}%
	\BibitemOpen
	\bibfield  {author} {\bibinfo {author} {\bibfnamefont {O.}~\bibnamefont
			{Morsch}}\ and\ \bibinfo {author} {\bibfnamefont {M.}~\bibnamefont
			{Oberthaler}},\ }\bibfield  {title} {\bibinfo {title} {Dynamics of
			{Bose}-{Einstein} condensates in optical lattices},\ }\href
	{https://doi.org/10.1103/RevModPhys.78.179} {\bibfield  {journal} {\bibinfo
			{journal} {Rev. Mod. Phys.}\ }\textbf {\bibinfo {volume} {78}},\ \bibinfo
		{pages} {179} (\bibinfo {year} {2006})}\BibitemShut {NoStop}%
	\bibitem [{\citenamefont {Dauxois}\ and\ \citenamefont
		{Peyrard}(2006)}]{dauxoisbook2006}%
	\BibitemOpen
	\bibfield  {author} {\bibinfo {author} {\bibfnamefont {T.}~\bibnamefont
			{Dauxois}}\ and\ \bibinfo {author} {\bibfnamefont {M.}~\bibnamefont
			{Peyrard}},\ }\href@noop {} {\emph {\bibinfo {title} {Physics of Solitons}}}\
	(\bibinfo  {publisher} {Cambridge University Press},\ \bibinfo {year}
	{2006})\BibitemShut {NoStop}%
	\bibitem [{\citenamefont {Kevrekidis}\ \emph {et~al.}(2008)\citenamefont
		{Kevrekidis}, \citenamefont {Frantzeskakis},\ and\ \citenamefont
		{Carretero-Gonz{\'a}lez}}]{BEC_book_2008}%
	\BibitemOpen
	\bibfield  {author} {\bibinfo {author} {\bibfnamefont {P.~G.}\ \bibnamefont
			{Kevrekidis}}, \bibinfo {author} {\bibfnamefont {D.~J.}\ \bibnamefont
			{Frantzeskakis}},\ and\ \bibinfo {author} {\bibfnamefont {R.}~\bibnamefont
			{Carretero-Gonz{\'a}lez}},\ }\href@noop {} {\emph {\bibinfo {title} {Emergent
				nonlinear phenomena in {Bose}-{Einstein} condensates: theory and
				experiment}}}\ (\bibinfo  {publisher} {Springer},\ \bibinfo {address}
	{Berlin, Heidelberg},\ \bibinfo {year} {2008})\BibitemShut {NoStop}%
	\bibitem [{\citenamefont {Frantzeskakis}(2010)}]{frantzeskakis2010dark}%
	\BibitemOpen
	\bibfield  {author} {\bibinfo {author} {\bibfnamefont {D.~J.}\ \bibnamefont
			{Frantzeskakis}},\ }\bibfield  {title} {\bibinfo {title} {Dark solitons in
			atomic bose--einstein condensates: from theory to experiments},\ }\href
	{https://doi.org/10.1088/1751-8113/43/21/213001} {\bibfield  {journal}
		{\bibinfo  {journal} {J. Phys. A}\ }\textbf {\bibinfo {volume} {43}},\
		\bibinfo {pages} {213001} (\bibinfo {year} {2010})}\BibitemShut {NoStop}%
	\bibitem [{\citenamefont {Liu}\ \emph {et~al.}(2018)\citenamefont {Liu},
		\citenamefont {Li}, \citenamefont {Fu},\ and\ \citenamefont
		{Ye}}]{liu2018nonlinear}%
	\BibitemOpen
	\bibfield  {author} {\bibinfo {author} {\bibfnamefont {J.}~\bibnamefont
			{Liu}}, \bibinfo {author} {\bibfnamefont {S.-C.}\ \bibnamefont {Li}},
		\bibinfo {author} {\bibfnamefont {L.-B.}\ \bibnamefont {Fu}},\ and\ \bibinfo
		{author} {\bibfnamefont {D.-F.}\ \bibnamefont {Ye}},\ }\href@noop {} {\emph
		{\bibinfo {title} {Nonlinear Adiabatic Evolution of Quantum Systems}}}\
	(\bibinfo  {publisher} {Springer},\ \bibinfo {year} {2018})\BibitemShut
	{NoStop}%
	\bibitem [{\citenamefont {Xu}\ \emph {et~al.}(2014)\citenamefont {Xu},
		\citenamefont {Mao}, \citenamefont {Wu},\ and\ \citenamefont
		{Zhang}}]{xu2014dark}%
	\BibitemOpen
	\bibfield  {author} {\bibinfo {author} {\bibfnamefont {Y.}~\bibnamefont
			{Xu}}, \bibinfo {author} {\bibfnamefont {L.}~\bibnamefont {Mao}}, \bibinfo
		{author} {\bibfnamefont {B.}~\bibnamefont {Wu}},\ and\ \bibinfo {author}
		{\bibfnamefont {C.}~\bibnamefont {Zhang}},\ }\bibfield  {title} {\bibinfo
		{title} {Dark solitons with majorana fermions in spin-orbit-coupled fermi
			gases},\ }\href {https://doi.org/10.1103/PhysRevLett.113.130404} {\bibfield
		{journal} {\bibinfo  {journal} {Phys. Rev. Lett.}\ }\textbf {\bibinfo
			{volume} {113}},\ \bibinfo {pages} {130404} (\bibinfo {year}
		{2014})}\BibitemShut {NoStop}%
	\bibitem [{\citenamefont {Liu}(2015)}]{liu2015soliton}%
	\BibitemOpen
	\bibfield  {author} {\bibinfo {author} {\bibfnamefont {X.-J.}\ \bibnamefont
			{Liu}},\ }\bibfield  {title} {\bibinfo {title} {Soliton-induced majorana
			fermions in a one-dimensional atomic topological superfluid},\ }\href
	{https://doi.org/10.1103/PhysRevA.91.023610} {\bibfield  {journal} {\bibinfo
			{journal} {Phys. Rev. A}\ }\textbf {\bibinfo {volume} {91}},\ \bibinfo
		{pages} {023610} (\bibinfo {year} {2015})}\BibitemShut {NoStop}%
	\bibitem [{\citenamefont {Zou}\ \emph {et~al.}(2016)\citenamefont {Zou},
		\citenamefont {Brand}, \citenamefont {Liu},\ and\ \citenamefont
		{Hu}}]{zou2016traveling}%
	\BibitemOpen
	\bibfield  {author} {\bibinfo {author} {\bibfnamefont {P.}~\bibnamefont
			{Zou}}, \bibinfo {author} {\bibfnamefont {J.}~\bibnamefont {Brand}}, \bibinfo
		{author} {\bibfnamefont {X.-J.}\ \bibnamefont {Liu}},\ and\ \bibinfo {author}
		{\bibfnamefont {H.}~\bibnamefont {Hu}},\ }\bibfield  {title} {\bibinfo
		{title} {Traveling majorana solitons in a low-dimensional spin-orbit-coupled
			fermi superfluid},\ }\href {https://doi.org/10.1103/PhysRevLett.117.225302}
	{\bibfield  {journal} {\bibinfo  {journal} {Phys. Rev. Lett.}\ }\textbf
		{\bibinfo {volume} {117}},\ \bibinfo {pages} {225302} (\bibinfo {year}
		{2016})}\BibitemShut {NoStop}%
	\bibitem [{\citenamefont {Zeng}\ \emph {et~al.}(2019)\citenamefont {Zeng},
		\citenamefont {Stanescu}, \citenamefont {Zhang}, \citenamefont {Scarola},\
		and\ \citenamefont {Tewari}}]{zeng2019majorana}%
	\BibitemOpen
	\bibfield  {author} {\bibinfo {author} {\bibfnamefont {C.}~\bibnamefont
			{Zeng}}, \bibinfo {author} {\bibfnamefont {T.~D.}\ \bibnamefont {Stanescu}},
		\bibinfo {author} {\bibfnamefont {C.}~\bibnamefont {Zhang}}, \bibinfo
		{author} {\bibfnamefont {V.~W.}\ \bibnamefont {Scarola}},\ and\ \bibinfo
		{author} {\bibfnamefont {S.}~\bibnamefont {Tewari}},\ }\bibfield  {title}
	{\bibinfo {title} {Majorana corner modes with solitons in an attractive
			hubbard-hofstadter model of cold atom optical lattices},\ }\href
	{https://doi.org/10.1103/PhysRevLett.123.060402} {\bibfield  {journal}
		{\bibinfo  {journal} {Phys. Rev. Lett.}\ }\textbf {\bibinfo {volume} {123}},\
		\bibinfo {pages} {060402} (\bibinfo {year} {2019})}\BibitemShut {NoStop}%
	\bibitem [{\citenamefont {Eisenberg}\ \emph {et~al.}(1998)\citenamefont
		{Eisenberg}, \citenamefont {Silberberg}, \citenamefont {Morandotti},
		\citenamefont {Boyd},\ and\ \citenamefont {Aitchison}}]{eisenbergPRL1998}%
	\BibitemOpen
	\bibfield  {author} {\bibinfo {author} {\bibfnamefont {H.~S.}\ \bibnamefont
			{Eisenberg}}, \bibinfo {author} {\bibfnamefont {Y.}~\bibnamefont
			{Silberberg}}, \bibinfo {author} {\bibfnamefont {R.}~\bibnamefont
			{Morandotti}}, \bibinfo {author} {\bibfnamefont {A.~R.}\ \bibnamefont
			{Boyd}},\ and\ \bibinfo {author} {\bibfnamefont {J.~S.}\ \bibnamefont
			{Aitchison}},\ }\bibfield  {title} {\bibinfo {title} {Discrete spatial
			optical solitons in waveguide arrays},\ }\href
	{https://doi.org/10.1103/PhysRevLett.81.3383} {\bibfield  {journal} {\bibinfo
			{journal} {Phys. Rev. Lett.}\ }\textbf {\bibinfo {volume} {81}},\ \bibinfo
		{pages} {3383} (\bibinfo {year} {1998})}\BibitemShut {NoStop}%
	\bibitem [{\citenamefont {Morandotti}\ \emph {et~al.}(2001)\citenamefont
		{Morandotti}, \citenamefont {Eisenberg}, \citenamefont {Silberberg},
		\citenamefont {Sorel},\ and\ \citenamefont {Aitchison}}]{morandotti2001self}%
	\BibitemOpen
	\bibfield  {author} {\bibinfo {author} {\bibfnamefont {R.}~\bibnamefont
			{Morandotti}}, \bibinfo {author} {\bibfnamefont {H.~S.}\ \bibnamefont
			{Eisenberg}}, \bibinfo {author} {\bibfnamefont {Y.}~\bibnamefont
			{Silberberg}}, \bibinfo {author} {\bibfnamefont {M.}~\bibnamefont {Sorel}},\
		and\ \bibinfo {author} {\bibfnamefont {J.~S.}\ \bibnamefont {Aitchison}},\
	}\bibfield  {title} {\bibinfo {title} {Self-focusing and defocusing in
			waveguide arrays},\ }\href {https://doi.org/10.1103/PhysRevLett.86.3296}
	{\bibfield  {journal} {\bibinfo  {journal} {Phys. Rev. Lett.}\ }\textbf
		{\bibinfo {volume} {86}},\ \bibinfo {pages} {3296} (\bibinfo {year}
		{2001})}\BibitemShut {NoStop}%
	\bibitem [{\citenamefont {Fleischer}\ \emph {et~al.}(2003)\citenamefont
		{Fleischer}, \citenamefont {Segev}, \citenamefont {Efremidis},\ and\
		\citenamefont {Christodoulides}}]{fleischerNat2003}%
	\BibitemOpen
	\bibfield  {author} {\bibinfo {author} {\bibfnamefont {J.~W.}\ \bibnamefont
			{Fleischer}}, \bibinfo {author} {\bibfnamefont {M.}~\bibnamefont {Segev}},
		\bibinfo {author} {\bibfnamefont {N.~K.}\ \bibnamefont {Efremidis}},\ and\
		\bibinfo {author} {\bibfnamefont {D.~N.}\ \bibnamefont {Christodoulides}},\
	}\bibfield  {title} {\bibinfo {title} {Observation of two-dimensional
			discrete solitons in optically induced nonlinear photonic lattices},\ }\href
	{https://doi.org/10.1038/nature01452} {\bibfield  {journal} {\bibinfo
			{journal} {Nature}\ }\textbf {\bibinfo {volume} {422}},\ \bibinfo {pages}
		{147} (\bibinfo {year} {2003})}\BibitemShut {NoStop}%
	\bibitem [{\citenamefont {Mandelik}\ \emph {et~al.}(2004)\citenamefont
		{Mandelik}, \citenamefont {Morandotti}, \citenamefont {Aitchison},\ and\
		\citenamefont {Silberberg}}]{mandelik2004gap}%
	\BibitemOpen
	\bibfield  {author} {\bibinfo {author} {\bibfnamefont {D.}~\bibnamefont
			{Mandelik}}, \bibinfo {author} {\bibfnamefont {R.}~\bibnamefont
			{Morandotti}}, \bibinfo {author} {\bibfnamefont {J.~S.}\ \bibnamefont
			{Aitchison}},\ and\ \bibinfo {author} {\bibfnamefont {Y.}~\bibnamefont
			{Silberberg}},\ }\bibfield  {title} {\bibinfo {title} {Gap solitons in
			waveguide arrays},\ }\href {https://doi.org/10.1103/PhysRevLett.92.093904}
	{\bibfield  {journal} {\bibinfo  {journal} {Phys. Rev. Lett.}\ }\textbf
		{\bibinfo {volume} {92}},\ \bibinfo {pages} {093904} (\bibinfo {year}
		{2004})}\BibitemShut {NoStop}%
	\bibitem [{\citenamefont {Burger}\ \emph {et~al.}(1999)\citenamefont {Burger},
		\citenamefont {Bongs}, \citenamefont {Dettmer}, \citenamefont {Ertmer},
		\citenamefont {Sengstock}, \citenamefont {Sanpera}, \citenamefont
		{Shlyapnikov},\ and\ \citenamefont {Lewenstein}}]{burger1999dark}%
	\BibitemOpen
	\bibfield  {author} {\bibinfo {author} {\bibfnamefont {S.}~\bibnamefont
			{Burger}}, \bibinfo {author} {\bibfnamefont {K.}~\bibnamefont {Bongs}},
		\bibinfo {author} {\bibfnamefont {S.}~\bibnamefont {Dettmer}}, \bibinfo
		{author} {\bibfnamefont {W.}~\bibnamefont {Ertmer}}, \bibinfo {author}
		{\bibfnamefont {K.}~\bibnamefont {Sengstock}}, \bibinfo {author}
		{\bibfnamefont {A.}~\bibnamefont {Sanpera}}, \bibinfo {author} {\bibfnamefont
			{G.~V.}\ \bibnamefont {Shlyapnikov}},\ and\ \bibinfo {author} {\bibfnamefont
			{M.}~\bibnamefont {Lewenstein}},\ }\bibfield  {title} {\bibinfo {title} {Dark
			solitons in bose-einstein condensates},\ }\href
	{https://doi.org/10.1103/PhysRevLett.83.5198} {\bibfield  {journal} {\bibinfo
			{journal} {Phys. Rev. Lett.}\ }\textbf {\bibinfo {volume} {83}},\ \bibinfo
		{pages} {5198} (\bibinfo {year} {1999})}\BibitemShut {NoStop}%
	\bibitem [{\citenamefont {Khaykovich}\ \emph {et~al.}(2002)\citenamefont
		{Khaykovich}, \citenamefont {Schreck}, \citenamefont {Ferrari}, \citenamefont
		{Bourdel}, \citenamefont {Cubizolles}, \citenamefont {Carr}, \citenamefont
		{Castin},\ and\ \citenamefont {Salomon}}]{khaykovich2002formation}%
	\BibitemOpen
	\bibfield  {author} {\bibinfo {author} {\bibfnamefont {L.}~\bibnamefont
			{Khaykovich}}, \bibinfo {author} {\bibfnamefont {F.}~\bibnamefont {Schreck}},
		\bibinfo {author} {\bibfnamefont {G.}~\bibnamefont {Ferrari}}, \bibinfo
		{author} {\bibfnamefont {T.}~\bibnamefont {Bourdel}}, \bibinfo {author}
		{\bibfnamefont {J.}~\bibnamefont {Cubizolles}}, \bibinfo {author}
		{\bibfnamefont {L.~D.}\ \bibnamefont {Carr}}, \bibinfo {author}
		{\bibfnamefont {Y.}~\bibnamefont {Castin}},\ and\ \bibinfo {author}
		{\bibfnamefont {C.}~\bibnamefont {Salomon}},\ }\bibfield  {title} {\bibinfo
		{title} {Formation of a matter-wave bright soliton},\ }\href
	{https://doi.org/10.1126/science.1071021} {\bibfield  {journal} {\bibinfo
			{journal} {Science}\ }\textbf {\bibinfo {volume} {296}},\ \bibinfo {pages}
		{1290} (\bibinfo {year} {2002})}\BibitemShut {NoStop}%
	\bibitem [{\citenamefont {Strecker}\ \emph {et~al.}(2002)\citenamefont
		{Strecker}, \citenamefont {Partridge}, \citenamefont {Truscott},\ and\
		\citenamefont {Hulet}}]{strecker2002formation}%
	\BibitemOpen
	\bibfield  {author} {\bibinfo {author} {\bibfnamefont {K.~E.}\ \bibnamefont
			{Strecker}}, \bibinfo {author} {\bibfnamefont {G.~B.}\ \bibnamefont
			{Partridge}}, \bibinfo {author} {\bibfnamefont {A.~G.}\ \bibnamefont
			{Truscott}},\ and\ \bibinfo {author} {\bibfnamefont {R.~G.}\ \bibnamefont
			{Hulet}},\ }\bibfield  {title} {\bibinfo {title} {Formation and propagation
			of matter-wave soliton trains},\ }\href {https://doi.org/10.1038/nature747}
	{\bibfield  {journal} {\bibinfo  {journal} {Nature}\ }\textbf {\bibinfo
			{volume} {417}},\ \bibinfo {pages} {150} (\bibinfo {year}
		{2002})}\BibitemShut {NoStop}%
	\bibitem [{\citenamefont {Klitzing}\ \emph {et~al.}(1980)\citenamefont
		{Klitzing}, \citenamefont {Dorda},\ and\ \citenamefont
		{Pepper}}]{klitzing1980new}%
	\BibitemOpen
	\bibfield  {author} {\bibinfo {author} {\bibfnamefont {K.~v.}\ \bibnamefont
			{Klitzing}}, \bibinfo {author} {\bibfnamefont {G.}~\bibnamefont {Dorda}},\
		and\ \bibinfo {author} {\bibfnamefont {M.}~\bibnamefont {Pepper}},\
	}\bibfield  {title} {\bibinfo {title} {New method for high-accuracy
			determination of the fine-structure constant based on quantized hall
			resistance},\ }\href {https://doi.org/10.1103/PhysRevLett.45.494} {\bibfield
		{journal} {\bibinfo  {journal} {Phys. Rev. Lett.}\ }\textbf {\bibinfo
			{volume} {45}},\ \bibinfo {pages} {494} (\bibinfo {year} {1980})}\BibitemShut
	{NoStop}%
	\bibitem [{\citenamefont {Thouless}\ \emph {et~al.}(1982)\citenamefont
		{Thouless}, \citenamefont {Kohmoto}, \citenamefont {Nightingale},\ and\
		\citenamefont {den Nijs}}]{thouless1982quantized}%
	\BibitemOpen
	\bibfield  {author} {\bibinfo {author} {\bibfnamefont {D.~J.}\ \bibnamefont
			{Thouless}}, \bibinfo {author} {\bibfnamefont {M.}~\bibnamefont {Kohmoto}},
		\bibinfo {author} {\bibfnamefont {M.~P.}\ \bibnamefont {Nightingale}},\ and\
		\bibinfo {author} {\bibfnamefont {M.}~\bibnamefont {den Nijs}},\ }\bibfield
	{title} {\bibinfo {title} {Quantized hall conductance in a two-dimensional
			periodic potential},\ }\href {https://doi.org/10.1103/PhysRevLett.49.405}
	{\bibfield  {journal} {\bibinfo  {journal} {Phys. Rev. Lett.}\ }\textbf
		{\bibinfo {volume} {49}},\ \bibinfo {pages} {405} (\bibinfo {year}
		{1982})}\BibitemShut {NoStop}%
	\bibitem [{\citenamefont {Simon}(1983)}]{simon1983holonomy}%
	\BibitemOpen
	\bibfield  {author} {\bibinfo {author} {\bibfnamefont {B.}~\bibnamefont
			{Simon}},\ }\bibfield  {title} {\bibinfo {title} {Holonomy, the quantum
			adiabatic theorem, and berry's phase},\ }\href
	{https://doi.org/10.1103/PhysRevLett.51.2167} {\bibfield  {journal} {\bibinfo
			{journal} {Phys. Rev. Lett.}\ }\textbf {\bibinfo {volume} {51}},\ \bibinfo
		{pages} {2167} (\bibinfo {year} {1983})}\BibitemShut {NoStop}%
	\bibitem [{\citenamefont {Laughlin}(1981)}]{laughlin1981quantized}%
	\BibitemOpen
	\bibfield  {author} {\bibinfo {author} {\bibfnamefont {R.~B.}\ \bibnamefont
			{Laughlin}},\ }\bibfield  {title} {\bibinfo {title} {Quantized hall
			conductivity in two dimensions},\ }\href
	{https://doi.org/10.1103/PhysRevB.23.5632} {\bibfield  {journal} {\bibinfo
			{journal} {Phys. Rev. B}\ }\textbf {\bibinfo {volume} {23}},\ \bibinfo
		{pages} {5632} (\bibinfo {year} {1981})}\BibitemShut {NoStop}%
	\bibitem [{\citenamefont {Thouless}(1983)}]{thoulessPRB1983}%
	\BibitemOpen
	\bibfield  {author} {\bibinfo {author} {\bibfnamefont {D.~J.}\ \bibnamefont
			{Thouless}},\ }\bibfield  {title} {\bibinfo {title} {Quantization of particle
			transport},\ }\href {https://doi.org/10.1103/PhysRevB.27.6083} {\bibfield
		{journal} {\bibinfo  {journal} {Phys. Rev. B}\ }\textbf {\bibinfo {volume}
			{27}},\ \bibinfo {pages} {6083} (\bibinfo {year} {1983})}\BibitemShut
	{NoStop}%
	\bibitem [{\citenamefont {Niu}\ and\ \citenamefont
		{Thouless}(1984)}]{niu1984quantised}%
	\BibitemOpen
	\bibfield  {author} {\bibinfo {author} {\bibfnamefont {Q.}~\bibnamefont
			{Niu}}\ and\ \bibinfo {author} {\bibfnamefont {D.~J.}\ \bibnamefont
			{Thouless}},\ }\bibfield  {title} {\bibinfo {title} {Quantised adiabatic
			charge transport in the presence of substrate disorder and many-body
			interaction},\ }\href {https://doi.org/10.1088/0305-4470/17/12/016}
	{\bibfield  {journal} {\bibinfo  {journal} {J. Phys. A Math. Gen.}\ }\textbf
		{\bibinfo {volume} {17}},\ \bibinfo {pages} {2453} (\bibinfo {year}
		{1984})}\BibitemShut {NoStop}%
	\bibitem [{\citenamefont {Ablowitz}\ \emph {et~al.}(2014)\citenamefont
		{Ablowitz}, \citenamefont {Curtis},\ and\ \citenamefont
		{Ma}}]{ablowitzPRA2014}%
	\BibitemOpen
	\bibfield  {author} {\bibinfo {author} {\bibfnamefont {M.~J.}\ \bibnamefont
			{Ablowitz}}, \bibinfo {author} {\bibfnamefont {C.~W.}\ \bibnamefont
			{Curtis}},\ and\ \bibinfo {author} {\bibfnamefont {Y.-P.}\ \bibnamefont
			{Ma}},\ }\bibfield  {title} {\bibinfo {title} {Linear and nonlinear traveling
			edge waves in optical honeycomb lattices},\ }\href
	{https://doi.org/10.1103/PhysRevA.90.023813} {\bibfield  {journal} {\bibinfo
			{journal} {Phys. Rev. A}\ }\textbf {\bibinfo {volume} {90}},\ \bibinfo
		{pages} {023813} (\bibinfo {year} {2014})}\BibitemShut {NoStop}%
	\bibitem [{\citenamefont {Leykam}\ and\ \citenamefont
		{Chong}(2016)}]{leykamPRL2016}%
	\BibitemOpen
	\bibfield  {author} {\bibinfo {author} {\bibfnamefont {D.}~\bibnamefont
			{Leykam}}\ and\ \bibinfo {author} {\bibfnamefont {Y.~D.}\ \bibnamefont
			{Chong}},\ }\bibfield  {title} {\bibinfo {title} {Edge solitons in
			nonlinear-photonic topological insulators},\ }\href
	{https://doi.org/10.1103/PhysRevLett.117.143901} {\bibfield  {journal}
		{\bibinfo  {journal} {Phys. Rev. Lett.}\ }\textbf {\bibinfo {volume} {117}},\
		\bibinfo {pages} {143901} (\bibinfo {year} {2016})}\BibitemShut {NoStop}%
	\bibitem [{\citenamefont {Mukherjee}\ and\ \citenamefont
		{Rechtsman}(2021)}]{mukherjeePRX2021}%
	\BibitemOpen
	\bibfield  {author} {\bibinfo {author} {\bibfnamefont {S.}~\bibnamefont
			{Mukherjee}}\ and\ \bibinfo {author} {\bibfnamefont {M.~C.}\ \bibnamefont
			{Rechtsman}},\ }\bibfield  {title} {\bibinfo {title} {Observation of
			unidirectional solitonlike edge states in nonlinear {Floquet} topological
			insulators},\ }\href {https://doi.org/10.1103/PhysRevX.11.041057} {\bibfield
		{journal} {\bibinfo  {journal} {Phys. Rev. X}\ }\textbf {\bibinfo {volume}
			{11}},\ \bibinfo {pages} {041057} (\bibinfo {year} {2021})}\BibitemShut
	{NoStop}%
	\bibitem [{\citenamefont {Tao}\ \emph {et~al.}(2020)\citenamefont {Tao},
		\citenamefont {Dai}, \citenamefont {Yang}, \citenamefont {Zeng},\ and\
		\citenamefont {Xu}}]{tao2020hinge}%
	\BibitemOpen
	\bibfield  {author} {\bibinfo {author} {\bibfnamefont {Y.-L.}\ \bibnamefont
			{Tao}}, \bibinfo {author} {\bibfnamefont {N.}~\bibnamefont {Dai}}, \bibinfo
		{author} {\bibfnamefont {Y.-B.}\ \bibnamefont {Yang}}, \bibinfo {author}
		{\bibfnamefont {Q.-B.}\ \bibnamefont {Zeng}},\ and\ \bibinfo {author}
		{\bibfnamefont {Y.}~\bibnamefont {Xu}},\ }\bibfield  {title} {\bibinfo
		{title} {Hinge solitons in three-dimensional second-order topological
			insulators},\ }\href {https://doi.org/10.1088/1367-2630/abc1f9} {\bibfield
		{journal} {\bibinfo  {journal} {New J. Phys.}\ }\textbf {\bibinfo {volume}
			{22}},\ \bibinfo {pages} {103058} (\bibinfo {year} {2020})}\BibitemShut
	{NoStop}%
	\bibitem [{\citenamefont {Maczewsky}\ \emph {et~al.}(2020)\citenamefont
		{Maczewsky}, \citenamefont {Heinrich}, \citenamefont {Kremer}, \citenamefont
		{Ivanov}, \citenamefont {Ehrhardt}, \citenamefont {Martinez}, \citenamefont
		{Kartashov}, \citenamefont {Konotop}, \citenamefont {Torner}, \citenamefont
		{Bauer},\ and\ \citenamefont {Szameit}}]{maczewskySci2020}%
	\BibitemOpen
	\bibfield  {author} {\bibinfo {author} {\bibfnamefont {L.~J.}\ \bibnamefont
			{Maczewsky}}, \bibinfo {author} {\bibfnamefont {M.}~\bibnamefont {Heinrich}},
		\bibinfo {author} {\bibfnamefont {M.}~\bibnamefont {Kremer}}, \bibinfo
		{author} {\bibfnamefont {S.~K.}\ \bibnamefont {Ivanov}}, \bibinfo {author}
		{\bibfnamefont {M.}~\bibnamefont {Ehrhardt}}, \bibinfo {author}
		{\bibfnamefont {F.}~\bibnamefont {Martinez}}, \bibinfo {author}
		{\bibfnamefont {Y.~V.}\ \bibnamefont {Kartashov}}, \bibinfo {author}
		{\bibfnamefont {V.~V.}\ \bibnamefont {Konotop}}, \bibinfo {author}
		{\bibfnamefont {L.}~\bibnamefont {Torner}}, \bibinfo {author} {\bibfnamefont
			{D.}~\bibnamefont {Bauer}},\ and\ \bibinfo {author} {\bibfnamefont
			{A.}~\bibnamefont {Szameit}},\ }\bibfield  {title} {\bibinfo {title}
		{Nonlinearity-induced photonic topological insulator},\ }\href
	{https://doi.org/10.1126/science.abd2033} {\bibfield  {journal} {\bibinfo
			{journal} {Science}\ }\textbf {\bibinfo {volume} {370}},\ \bibinfo {pages}
		{701} (\bibinfo {year} {2020})}\BibitemShut {NoStop}%
	\bibitem [{\citenamefont {Sone}\ \emph {et~al.}(2024)\citenamefont {Sone},
		\citenamefont {Ezawa}, \citenamefont {Ashida}, \citenamefont {Yoshioka},\
		and\ \citenamefont {Sagawa}}]{soneNP2024}%
	\BibitemOpen
	\bibfield  {author} {\bibinfo {author} {\bibfnamefont {K.}~\bibnamefont
			{Sone}}, \bibinfo {author} {\bibfnamefont {M.}~\bibnamefont {Ezawa}},
		\bibinfo {author} {\bibfnamefont {Y.}~\bibnamefont {Ashida}}, \bibinfo
		{author} {\bibfnamefont {N.}~\bibnamefont {Yoshioka}},\ and\ \bibinfo
		{author} {\bibfnamefont {T.}~\bibnamefont {Sagawa}},\ }\bibfield  {title}
	{\bibinfo {title} {Nonlinearity-induced topological phase transition
			characterized by the nonlinear {Chern} number},\ }\href
	{https://doi.org/10.1038/s41567-024-02451-x} {\bibfield  {journal} {\bibinfo
			{journal} {Nat. Phys.}\ }\textbf {\bibinfo {volume} {20}},\ \bibinfo {pages}
		{1164} (\bibinfo {year} {2024})}\BibitemShut {NoStop}%
	\bibitem [{\citenamefont {J{\"u}rgensen}\ \emph {et~al.}(2021)\citenamefont
		{J{\"u}rgensen}, \citenamefont {Mukherjee},\ and\ \citenamefont
		{Rechtsman}}]{jurgensenNat2021}%
	\BibitemOpen
	\bibfield  {author} {\bibinfo {author} {\bibfnamefont {M.}~\bibnamefont
			{J{\"u}rgensen}}, \bibinfo {author} {\bibfnamefont {S.}~\bibnamefont
			{Mukherjee}},\ and\ \bibinfo {author} {\bibfnamefont {M.~C.}\ \bibnamefont
			{Rechtsman}},\ }\bibfield  {title} {\bibinfo {title} {Quantized nonlinear
			{Thouless} pumping},\ }\href {https://doi.org/10.1038/s41586-021-03688-9}
	{\bibfield  {journal} {\bibinfo  {journal} {Nature}\ }\textbf {\bibinfo
			{volume} {596}},\ \bibinfo {pages} {63} (\bibinfo {year} {2021})}\BibitemShut
	{NoStop}%
	\bibitem [{\citenamefont {J{\"u}rgensen}\ and\ \citenamefont
		{Rechtsman}(2022)}]{jurgensenPRL2022}%
	\BibitemOpen
	\bibfield  {author} {\bibinfo {author} {\bibfnamefont {M.}~\bibnamefont
			{J{\"u}rgensen}}\ and\ \bibinfo {author} {\bibfnamefont {M.~C.}\ \bibnamefont
			{Rechtsman}},\ }\bibfield  {title} {\bibinfo {title} {Chern number governs
			soliton motion in nonlinear thouless pumps},\ }\href
	{https://doi.org/10.1103/PhysRevLett.128.113901} {\bibfield  {journal}
		{\bibinfo  {journal} {Phys. Rev. Lett.}\ }\textbf {\bibinfo {volume} {128}},\
		\bibinfo {pages} {113901} (\bibinfo {year} {2022})}\BibitemShut {NoStop}%
	\bibitem [{\citenamefont {Fu}\ \emph {et~al.}(2022{\natexlab{a}})\citenamefont
		{Fu}, \citenamefont {Wang}, \citenamefont {Kartashov}, \citenamefont
		{Konotop},\ and\ \citenamefont {Ye}}]{fuPRL2022}%
	\BibitemOpen
	\bibfield  {author} {\bibinfo {author} {\bibfnamefont {Q.}~\bibnamefont
			{Fu}}, \bibinfo {author} {\bibfnamefont {P.}~\bibnamefont {Wang}}, \bibinfo
		{author} {\bibfnamefont {Y.~V.}\ \bibnamefont {Kartashov}}, \bibinfo {author}
		{\bibfnamefont {V.~V.}\ \bibnamefont {Konotop}},\ and\ \bibinfo {author}
		{\bibfnamefont {F.}~\bibnamefont {Ye}},\ }\bibfield  {title} {\bibinfo
		{title} {Nonlinear {Thouless} pumping: solitons and transport breakdown},\
	}\href {https://doi.org/10.1103/PhysRevLett.128.154101} {\bibfield  {journal}
		{\bibinfo  {journal} {Phys. Rev. Lett.}\ }\textbf {\bibinfo {volume} {128}},\
		\bibinfo {pages} {154101} (\bibinfo {year} {2022}{\natexlab{a}})}\BibitemShut
	{NoStop}%
	\bibitem [{\citenamefont {Fu}\ \emph {et~al.}(2022{\natexlab{b}})\citenamefont
		{Fu}, \citenamefont {Wang}, \citenamefont {Kartashov}, \citenamefont
		{Konotop},\ and\ \citenamefont {Ye}}]{fu2022twoD}%
	\BibitemOpen
	\bibfield  {author} {\bibinfo {author} {\bibfnamefont {Q.}~\bibnamefont
			{Fu}}, \bibinfo {author} {\bibfnamefont {P.}~\bibnamefont {Wang}}, \bibinfo
		{author} {\bibfnamefont {Y.~V.}\ \bibnamefont {Kartashov}}, \bibinfo {author}
		{\bibfnamefont {V.~V.}\ \bibnamefont {Konotop}},\ and\ \bibinfo {author}
		{\bibfnamefont {F.}~\bibnamefont {Ye}},\ }\bibfield  {title} {\bibinfo
		{title} {Two-dimensional nonlinear {Thouless} pumping of matter waves},\
	}\href {https://doi.org/10.1103/PhysRevLett.129.183901} {\bibfield  {journal}
		{\bibinfo  {journal} {Phys. Rev. Lett.}\ }\textbf {\bibinfo {volume} {129}},\
		\bibinfo {pages} {183901} (\bibinfo {year} {2022}{\natexlab{b}})}\BibitemShut
	{NoStop}%
	\bibitem [{\citenamefont {Mostaan}\ \emph {et~al.}(2022)\citenamefont
		{Mostaan}, \citenamefont {Grusdt},\ and\ \citenamefont
		{Goldman}}]{mostaanNC2022}%
	\BibitemOpen
	\bibfield  {author} {\bibinfo {author} {\bibfnamefont {N.}~\bibnamefont
			{Mostaan}}, \bibinfo {author} {\bibfnamefont {F.}~\bibnamefont {Grusdt}},\
		and\ \bibinfo {author} {\bibfnamefont {N.}~\bibnamefont {Goldman}},\
	}\bibfield  {title} {\bibinfo {title} {Quantized topological pumping of
			solitons in nonlinear photonics and ultracold atomic mixtures},\ }\href
	{https://doi.org/10.1038/s41467-022-33478-4} {\bibfield  {journal} {\bibinfo
			{journal} {Nat. Commun.}\ }\textbf {\bibinfo {volume} {13}},\ \bibinfo
		{pages} {5997} (\bibinfo {year} {2022})}\BibitemShut {NoStop}%
	\bibitem [{\citenamefont {Tuloup}\ \emph {et~al.}(2023)\citenamefont {Tuloup},
		\citenamefont {Bomantara},\ and\ \citenamefont {Gong}}]{tuloupNJP2023}%
	\BibitemOpen
	\bibfield  {author} {\bibinfo {author} {\bibfnamefont {T.}~\bibnamefont
			{Tuloup}}, \bibinfo {author} {\bibfnamefont {R.~W.}\ \bibnamefont
			{Bomantara}},\ and\ \bibinfo {author} {\bibfnamefont {J.}~\bibnamefont
			{Gong}},\ }\bibfield  {title} {\bibinfo {title} {Breakdown of quantization in
			nonlinear {Thouless} pumping},\ }\href
	{https://doi.org/10.1088/1367-2630/acef4d} {\bibfield  {journal} {\bibinfo
			{journal} {New J. Phys.}\ }\textbf {\bibinfo {volume} {25}},\ \bibinfo
		{pages} {083048} (\bibinfo {year} {2023})}\BibitemShut {NoStop}%
	\bibitem [{\citenamefont {Citro}\ and\ \citenamefont
		{Aidelsburger}(2023)}]{citro2023thouless}%
	\BibitemOpen
	\bibfield  {author} {\bibinfo {author} {\bibfnamefont {R.}~\bibnamefont
			{Citro}}\ and\ \bibinfo {author} {\bibfnamefont {M.}~\bibnamefont
			{Aidelsburger}},\ }\bibfield  {title} {\bibinfo {title} {Thouless pumping and
			topology},\ }\href {https://doi.org/10.1038/s42254-022-00545-0} {\bibfield
		{journal} {\bibinfo  {journal} {Nat. Rev. Phys.}\ }\textbf {\bibinfo {volume}
			{5}},\ \bibinfo {pages} {87} (\bibinfo {year} {2023})}\BibitemShut {NoStop}%
	\bibitem [{\citenamefont {Hu}\ \emph {et~al.}(2024)\citenamefont {Hu},
		\citenamefont {Li}, \citenamefont {Chen},\ and\ \citenamefont
		{Luo}}]{HuNJP2024}%
	\BibitemOpen
	\bibfield  {author} {\bibinfo {author} {\bibfnamefont {X.}~\bibnamefont
			{Hu}}, \bibinfo {author} {\bibfnamefont {Z.}~\bibnamefont {Li}}, \bibinfo
		{author} {\bibfnamefont {A.-X.}\ \bibnamefont {Chen}},\ and\ \bibinfo
		{author} {\bibfnamefont {X.}~\bibnamefont {Luo}},\ }\bibfield  {title}
	{\bibinfo {title} {Pumping of matter wave solitons in one-dimensional optical
			superlattices},\ }\href {https://doi.org/10.1088/1367-2630/ad9770} {\bibfield
		{journal} {\bibinfo  {journal} {New J. Phys.}\ }\textbf {\bibinfo {volume}
			{26}},\ \bibinfo {pages} {123006} (\bibinfo {year} {2024})}\BibitemShut
	{NoStop}%
	\bibitem [{\citenamefont {Cao}\ \emph {et~al.}(2024)\citenamefont {Cao},
		\citenamefont {Jia}, \citenamefont {Hu},\ and\ \citenamefont
		{Liang}}]{cao2024nonlinear}%
	\BibitemOpen
	\bibfield  {author} {\bibinfo {author} {\bibfnamefont {X.}~\bibnamefont
			{Cao}}, \bibinfo {author} {\bibfnamefont {C.}~\bibnamefont {Jia}}, \bibinfo
		{author} {\bibfnamefont {Y.}~\bibnamefont {Hu}},\ and\ \bibinfo {author}
		{\bibfnamefont {Z.}~\bibnamefont {Liang}},\ }\bibfield  {title} {\bibinfo
		{title} {Nonlinear thouless pumping of solitons across an impurity},\ }\href
	{https://doi.org/10.1103/PhysRevA.110.013305} {\bibfield  {journal} {\bibinfo
			{journal} {Phys. Rev. A}\ }\textbf {\bibinfo {volume} {110}},\ \bibinfo
		{pages} {013305} (\bibinfo {year} {2024})}\BibitemShut {NoStop}%
	\bibitem [{\citenamefont {Lyu}\ \emph {et~al.}(2024)\citenamefont {Lyu},
		\citenamefont {Zhang},\ and\ \citenamefont {Busch}}]{lyuPRR2024}%
	\BibitemOpen
	\bibfield  {author} {\bibinfo {author} {\bibfnamefont {H.}~\bibnamefont
			{Lyu}}, \bibinfo {author} {\bibfnamefont {Y.}~\bibnamefont {Zhang}},\ and\
		\bibinfo {author} {\bibfnamefont {T.}~\bibnamefont {Busch}},\ }\bibfield
	{title} {\bibinfo {title} {Thouless pumping and trapping of two-component gap
			solitons},\ }\href {https://doi.org/10.1103/PhysRevResearch.6.L042010}
	{\bibfield  {journal} {\bibinfo  {journal} {Phys. Rev. Research}\ }\textbf
		{\bibinfo {volume} {6}},\ \bibinfo {pages} {L042010} (\bibinfo {year}
		{2024})}\BibitemShut {NoStop}%
	\bibitem [{\citenamefont {Szameit}\ and\ \citenamefont
		{Rechtsman}(2024)}]{szameit2024discrete}%
	\BibitemOpen
	\bibfield  {author} {\bibinfo {author} {\bibfnamefont {A.}~\bibnamefont
			{Szameit}}\ and\ \bibinfo {author} {\bibfnamefont {M.~C.}\ \bibnamefont
			{Rechtsman}},\ }\bibfield  {title} {\bibinfo {title} {Discrete nonlinear
			topological photonics},\ }\href {https://doi.org/10.1038/s41567-024-02454-8}
	{\bibfield  {journal} {\bibinfo  {journal} {Nat. Phys.}\ }\textbf {\bibinfo
			{volume} {20}},\ \bibinfo {pages} {905} (\bibinfo {year} {2024})}\BibitemShut
	{NoStop}%
	\bibitem [{\citenamefont {Cao}\ \emph {et~al.}(2025)\citenamefont {Cao},
		\citenamefont {Jia}, \citenamefont {Lyu}, \citenamefont {Hu},\ and\
		\citenamefont {Liang}}]{cao2025transport}%
	\BibitemOpen
	\bibfield  {author} {\bibinfo {author} {\bibfnamefont {X.}~\bibnamefont
			{Cao}}, \bibinfo {author} {\bibfnamefont {C.}~\bibnamefont {Jia}}, \bibinfo
		{author} {\bibfnamefont {H.}~\bibnamefont {Lyu}}, \bibinfo {author}
		{\bibfnamefont {Y.}~\bibnamefont {Hu}},\ and\ \bibinfo {author}
		{\bibfnamefont {Z.}~\bibnamefont {Liang}},\ }\bibfield  {title} {\bibinfo
		{title} {Transport of vector solitons in spin-dependent nonlinear thouless
			pumps},\ }\href {https://doi.org/10.1103/PhysRevA.111.023329} {\bibfield
		{journal} {\bibinfo  {journal} {Phys. Rev. A}\ }\textbf {\bibinfo {volume}
			{111}},\ \bibinfo {pages} {023329} (\bibinfo {year} {2025})}\BibitemShut
	{NoStop}%
	\bibitem [{\citenamefont {Muryshev}\ \emph {et~al.}(1999)\citenamefont
		{Muryshev}, \citenamefont {van~den Heuvell},\ and\ \citenamefont
		{Shlyapnikov}}]{muryshev1999stability}%
	\BibitemOpen
	\bibfield  {author} {\bibinfo {author} {\bibfnamefont {A.~E.}\ \bibnamefont
			{Muryshev}}, \bibinfo {author} {\bibfnamefont {H.~B. v.~L.}\ \bibnamefont
			{van~den Heuvell}},\ and\ \bibinfo {author} {\bibfnamefont {G.~V.}\
			\bibnamefont {Shlyapnikov}},\ }\bibfield  {title} {\bibinfo {title}
		{Stability of standing matter waves in a trap},\ }\href
	{https://doi.org/10.1103/PhysRevA.60.R2665} {\bibfield  {journal} {\bibinfo
			{journal} {Phys. Rev. A}\ }\textbf {\bibinfo {volume} {60}},\ \bibinfo
		{pages} {R2665} (\bibinfo {year} {1999})}\BibitemShut {NoStop}%
	\bibitem [{\citenamefont {Fedichev}\ \emph {et~al.}(1999)\citenamefont
		{Fedichev}, \citenamefont {Muryshev},\ and\ \citenamefont
		{Shlyapnikov}}]{fedichev1999dissipative}%
	\BibitemOpen
	\bibfield  {author} {\bibinfo {author} {\bibfnamefont {P.~O.}\ \bibnamefont
			{Fedichev}}, \bibinfo {author} {\bibfnamefont {A.~E.}\ \bibnamefont
			{Muryshev}},\ and\ \bibinfo {author} {\bibfnamefont {G.~V.}\ \bibnamefont
			{Shlyapnikov}},\ }\bibfield  {title} {\bibinfo {title} {Dissipative dynamics
			of a kink state in a bose-condensed gas},\ }\href
	{https://doi.org/10.1103/PhysRevA.60.3220} {\bibfield  {journal} {\bibinfo
			{journal} {Phys. Rev. A}\ }\textbf {\bibinfo {volume} {60}},\ \bibinfo
		{pages} {3220} (\bibinfo {year} {1999})}\BibitemShut {NoStop}%
	\bibitem [{\citenamefont {Pelinovsky}\ and\ \citenamefont
		{Kevrekidis}(2008)}]{pelinovsky2008stability}%
	\BibitemOpen
	\bibfield  {author} {\bibinfo {author} {\bibfnamefont {D.~E.}\ \bibnamefont
			{Pelinovsky}}\ and\ \bibinfo {author} {\bibfnamefont {P.~G.}\ \bibnamefont
			{Kevrekidis}},\ }\bibfield  {title} {\bibinfo {title} {Stability of discrete
			dark solitons in nonlinear schr{\"o}dinger lattices},\ }\href
	{https://doi.org/10.1088/1751-8113/41/18/185206} {\bibfield  {journal}
		{\bibinfo  {journal} {J. Phys. A: Math. Theor.}\ }\textbf {\bibinfo {volume}
			{41}},\ \bibinfo {pages} {185206} (\bibinfo {year} {2008})}\BibitemShut
	{NoStop}%
	\bibitem [{\citenamefont {Wang}\ \emph {et~al.}(2013)\citenamefont {Wang},
		\citenamefont {Troyer},\ and\ \citenamefont {Dai}}]{wang2013topological}%
	\BibitemOpen
	\bibfield  {author} {\bibinfo {author} {\bibfnamefont {L.}~\bibnamefont
			{Wang}}, \bibinfo {author} {\bibfnamefont {M.}~\bibnamefont {Troyer}},\ and\
		\bibinfo {author} {\bibfnamefont {X.}~\bibnamefont {Dai}},\ }\bibfield
	{title} {\bibinfo {title} {Topological charge pumping in a one-dimensional
			optical lattice},\ }\href {https://doi.org/10.1103/PhysRevLett.111.026802}
	{\bibfield  {journal} {\bibinfo  {journal} {Phys. Rev. Lett.}\ }\textbf
		{\bibinfo {volume} {111}},\ \bibinfo {pages} {026802} (\bibinfo {year}
		{2013})}\BibitemShut {NoStop}%
	\bibitem [{\citenamefont {Nakajima}\ \emph {et~al.}(2016)\citenamefont
		{Nakajima}, \citenamefont {Tomita}, \citenamefont {Taie}, \citenamefont
		{Ichinose}, \citenamefont {Ozawa}, \citenamefont {Wang}, \citenamefont
		{Troyer},\ and\ \citenamefont {Takahashi}}]{nakajima2016topological}%
	\BibitemOpen
	\bibfield  {author} {\bibinfo {author} {\bibfnamefont {S.}~\bibnamefont
			{Nakajima}}, \bibinfo {author} {\bibfnamefont {T.}~\bibnamefont {Tomita}},
		\bibinfo {author} {\bibfnamefont {S.}~\bibnamefont {Taie}}, \bibinfo {author}
		{\bibfnamefont {T.}~\bibnamefont {Ichinose}}, \bibinfo {author}
		{\bibfnamefont {H.}~\bibnamefont {Ozawa}}, \bibinfo {author} {\bibfnamefont
			{L.}~\bibnamefont {Wang}}, \bibinfo {author} {\bibfnamefont {M.}~\bibnamefont
			{Troyer}},\ and\ \bibinfo {author} {\bibfnamefont {Y.}~\bibnamefont
			{Takahashi}},\ }\bibfield  {title} {\bibinfo {title} {Topological thouless
			pumping of ultracold fermions},\ }\href {https://doi.org/10.1038/nphys3622}
	{\bibfield  {journal} {\bibinfo  {journal} {Nat. Phys.}\ }\textbf {\bibinfo
			{volume} {12}},\ \bibinfo {pages} {296} (\bibinfo {year} {2016})}\BibitemShut
	{NoStop}%
	\bibitem [{\citenamefont {Lohse}\ \emph {et~al.}(2016)\citenamefont {Lohse},
		\citenamefont {Schweizer}, \citenamefont {Zilberberg}, \citenamefont
		{Aidelsburger},\ and\ \citenamefont {Bloch}}]{lohse2016thouless}%
	\BibitemOpen
	\bibfield  {author} {\bibinfo {author} {\bibfnamefont {M.}~\bibnamefont
			{Lohse}}, \bibinfo {author} {\bibfnamefont {C.}~\bibnamefont {Schweizer}},
		\bibinfo {author} {\bibfnamefont {O.}~\bibnamefont {Zilberberg}}, \bibinfo
		{author} {\bibfnamefont {M.}~\bibnamefont {Aidelsburger}},\ and\ \bibinfo
		{author} {\bibfnamefont {I.}~\bibnamefont {Bloch}},\ }\bibfield  {title}
	{\bibinfo {title} {A thouless quantum pump with ultracold bosonic atoms in an
			optical superlattice},\ }\href {https://doi.org/10.1038/nphys3584} {\bibfield
		{journal} {\bibinfo  {journal} {Nat. Phys.}\ }\textbf {\bibinfo {volume}
			{12}},\ \bibinfo {pages} {350} (\bibinfo {year} {2016})}\BibitemShut
	{NoStop}%
	\bibitem [{\citenamefont {Xiao}\ \emph {et~al.}(2010)\citenamefont {Xiao},
		\citenamefont {Chang},\ and\ \citenamefont {Niu}}]{xiao2010berry}%
	\BibitemOpen
	\bibfield  {author} {\bibinfo {author} {\bibfnamefont {D.}~\bibnamefont
			{Xiao}}, \bibinfo {author} {\bibfnamefont {M.-C.}\ \bibnamefont {Chang}},\
		and\ \bibinfo {author} {\bibfnamefont {Q.}~\bibnamefont {Niu}},\ }\bibfield
	{title} {\bibinfo {title} {Berry phase effects on electronic properties},\
	}\href {https://doi.org/10.1103/RevModPhys.82.1959} {\bibfield  {journal}
		{\bibinfo  {journal} {Rev. Mod. Phys.}\ }\textbf {\bibinfo {volume} {82}},\
		\bibinfo {pages} {1959} (\bibinfo {year} {2010})}\BibitemShut {NoStop}%
	\bibitem [{\citenamefont {Kivshar}\ and\ \citenamefont
		{Malomed}(1989)}]{kivsharRMP1989}%
	\BibitemOpen
	\bibfield  {author} {\bibinfo {author} {\bibfnamefont {Y.~S.}\ \bibnamefont
			{Kivshar}}\ and\ \bibinfo {author} {\bibfnamefont {B.~A.}\ \bibnamefont
			{Malomed}},\ }\bibfield  {title} {\bibinfo {title} {Dynamics of solitons in
			nearly integrable systems},\ }\href
	{https://doi.org/10.1103/revmodphys.61.763} {\bibfield  {journal} {\bibinfo
			{journal} {Rev. Mod. Phys.}\ }\textbf {\bibinfo {volume} {61}},\ \bibinfo
		{pages} {763} (\bibinfo {year} {1989})}\BibitemShut {NoStop}%
	\bibitem [{\citenamefont {Band}\ and\ \citenamefont
		{Trippenbach}(2002)}]{bandPRA2002}%
	\BibitemOpen
	\bibfield  {author} {\bibinfo {author} {\bibfnamefont {Y.~B.}\ \bibnamefont
			{Band}}\ and\ \bibinfo {author} {\bibfnamefont {M.}~\bibnamefont
			{Trippenbach}},\ }\bibfield  {title} {\bibinfo {title} {Bose-{Einstein}
			condensates in time-dependent light potentials: {Adiabatic} and nonadiabatic
			behavior of nonlinear wave equations},\ }\href
	{https://doi.org/10.1103/physreva.65.053602} {\bibfield  {journal} {\bibinfo
			{journal} {Phys. Rev. A}\ }\textbf {\bibinfo {volume} {65}},\ \bibinfo
		{pages} {053602} (\bibinfo {year} {2002})}\BibitemShut {NoStop}%
	\bibitem [{\citenamefont {Band}\ \emph {et~al.}(2002)\citenamefont {Band},
		\citenamefont {Malomed},\ and\ \citenamefont {Trippenbach}}]{bandPRA2002_2}%
	\BibitemOpen
	\bibfield  {author} {\bibinfo {author} {\bibfnamefont {Y.~B.}\ \bibnamefont
			{Band}}, \bibinfo {author} {\bibfnamefont {B.}~\bibnamefont {Malomed}},\ and\
		\bibinfo {author} {\bibfnamefont {M.}~\bibnamefont {Trippenbach}},\
	}\bibfield  {title} {\bibinfo {title} {Adiabaticity in nonlinear quantum
			dynamics: {Bose}-{Einstein} condensate in a time-varying box},\ }\href
	{https://doi.org/10.1103/physreva.65.033607} {\bibfield  {journal} {\bibinfo
			{journal} {Phys. Rev. A}\ }\textbf {\bibinfo {volume} {65}},\ \bibinfo
		{pages} {033607} (\bibinfo {year} {2002})}\BibitemShut {NoStop}%
	\bibitem [{\citenamefont {Liu}\ \emph {et~al.}(2003)\citenamefont {Liu},
		\citenamefont {Wu},\ and\ \citenamefont {Niu}}]{liuPRL2003}%
	\BibitemOpen
	\bibfield  {author} {\bibinfo {author} {\bibfnamefont {J.}~\bibnamefont
			{Liu}}, \bibinfo {author} {\bibfnamefont {B.}~\bibnamefont {Wu}},\ and\
		\bibinfo {author} {\bibfnamefont {Q.}~\bibnamefont {Niu}},\ }\bibfield
	{title} {\bibinfo {title} {Nonlinear evolution of quantum states in the
			adiabatic regime},\ }\href {https://doi.org/10.1103/physrevlett.90.170404}
	{\bibfield  {journal} {\bibinfo  {journal} {Phys. Rev. Lett.}\ }\textbf
		{\bibinfo {volume} {90}},\ \bibinfo {pages} {170404} (\bibinfo {year}
		{2003})}\BibitemShut {NoStop}%
	\bibitem [{\citenamefont {Wu}\ \emph {et~al.}(2005)\citenamefont {Wu},
		\citenamefont {Liu},\ and\ \citenamefont {Niu}}]{wuPRL2005}%
	\BibitemOpen
	\bibfield  {author} {\bibinfo {author} {\bibfnamefont {B.}~\bibnamefont
			{Wu}}, \bibinfo {author} {\bibfnamefont {J.}~\bibnamefont {Liu}},\ and\
		\bibinfo {author} {\bibfnamefont {Q.}~\bibnamefont {Niu}},\ }\bibfield
	{title} {\bibinfo {title} {Geometric phase for adiabatic evolutions of
			general quantum states},\ }\href
	{https://doi.org/10.1103/PhysRevLett.94.140402} {\bibfield  {journal}
		{\bibinfo  {journal} {Phys. Rev. Lett.}\ }\textbf {\bibinfo {volume} {94}},\
		\bibinfo {pages} {140402} (\bibinfo {year} {2005})}\BibitemShut {NoStop}%
	\bibitem [{\citenamefont {Parker}\ \emph {et~al.}(2003)\citenamefont {Parker},
		\citenamefont {Proukakis}, \citenamefont {Leadbeater},\ and\ \citenamefont
		{Adams}}]{parker2003deformation}%
	\BibitemOpen
	\bibfield  {author} {\bibinfo {author} {\bibfnamefont {N.~G.}\ \bibnamefont
			{Parker}}, \bibinfo {author} {\bibfnamefont {N.~P.}\ \bibnamefont
			{Proukakis}}, \bibinfo {author} {\bibfnamefont {M.}~\bibnamefont
			{Leadbeater}},\ and\ \bibinfo {author} {\bibfnamefont {C.~S.}\ \bibnamefont
			{Adams}},\ }\bibfield  {title} {\bibinfo {title} {Deformation of dark
			solitons in inhomogeneous bose--einstein condensates},\ }\href
	{https://doi.org/10.1088/0953-4075/36/13/318} {\bibfield  {journal} {\bibinfo
			{journal} {J. Phys. B: At. Mol. Opt. Phys.}\ }\textbf {\bibinfo {volume}
			{36}},\ \bibinfo {pages} {2891} (\bibinfo {year} {2003})}\BibitemShut
	{NoStop}%
	\bibitem [{\citenamefont {J{\"u}rgensen}\ \emph {et~al.}(2023)\citenamefont
		{J{\"u}rgensen}, \citenamefont {Mukherjee}, \citenamefont {J{\"o}rg},\ and\
		\citenamefont {Rechtsman}}]{jurgensenNP2023}%
	\BibitemOpen
	\bibfield  {author} {\bibinfo {author} {\bibfnamefont {M.}~\bibnamefont
			{J{\"u}rgensen}}, \bibinfo {author} {\bibfnamefont {S.}~\bibnamefont
			{Mukherjee}}, \bibinfo {author} {\bibfnamefont {C.}~\bibnamefont
			{J{\"o}rg}},\ and\ \bibinfo {author} {\bibfnamefont {M.~C.}\ \bibnamefont
			{Rechtsman}},\ }\bibfield  {title} {\bibinfo {title} {Quantized fractional
			thouless pumping of solitons},\ }\href
	{https://doi.org/10.1038/s41567-022-01871-x} {\bibfield  {journal} {\bibinfo
			{journal} {Nat. Phys.}\ }\textbf {\bibinfo {volume} {19}},\ \bibinfo {pages}
		{420} (\bibinfo {year} {2023})}\BibitemShut {NoStop}%
	\bibitem [{\citenamefont {Perez-Garcia}\ \emph {et~al.}(1998)\citenamefont
		{Perez-Garcia}, \citenamefont {Michinel},\ and\ \citenamefont
		{Herrero}}]{perez1998bose}%
	\BibitemOpen
	\bibfield  {author} {\bibinfo {author} {\bibfnamefont {V.~M.}\ \bibnamefont
			{Perez-Garcia}}, \bibinfo {author} {\bibfnamefont {H.}~\bibnamefont
			{Michinel}},\ and\ \bibinfo {author} {\bibfnamefont {H.}~\bibnamefont
			{Herrero}},\ }\bibfield  {title} {\bibinfo {title} {Bose-einstein solitons in
			highly asymmetric traps},\ }\href {https://doi.org/10.1103/PhysRevA.57.3837}
	{\bibfield  {journal} {\bibinfo  {journal} {Phys. Rev. A}\ }\textbf {\bibinfo
			{volume} {57}},\ \bibinfo {pages} {3837} (\bibinfo {year}
		{1998})}\BibitemShut {NoStop}%
	\bibitem [{\citenamefont {Tao}\ \emph {et~al.}(2024)\citenamefont {Tao},
		\citenamefont {Wang},\ and\ \citenamefont {Xu}}]{tao2024nonlinearity}%
	\BibitemOpen
	\bibfield  {author} {\bibinfo {author} {\bibfnamefont {Y.-L.}\ \bibnamefont
			{Tao}}, \bibinfo {author} {\bibfnamefont {J.-H.}\ \bibnamefont {Wang}},\ and\
		\bibinfo {author} {\bibfnamefont {Y.}~\bibnamefont {Xu}},\ }\href@noop {}
	{\bibinfo {title} {Nonlinearity-induced thouless pumping of solitons}}
	(\bibinfo {year} {2024}),\ \Eprint {https://arxiv.org/abs/2409.19515}
	{arXiv:2409.19515} \BibitemShut {NoStop}%
	\bibitem [{\citenamefont {Tao}\ \emph {et~al.}(2025)\citenamefont {Tao},
		\citenamefont {Zhang},\ and\ \citenamefont {Xu}}]{tao2024nonlinearity_frac}%
	\BibitemOpen
	\bibfield  {author} {\bibinfo {author} {\bibfnamefont {Y.-L.}\ \bibnamefont
			{Tao}}, \bibinfo {author} {\bibfnamefont {Y.}~\bibnamefont {Zhang}},\ and\
		\bibinfo {author} {\bibfnamefont {Y.}~\bibnamefont {Xu}},\ }\href@noop {}
	{\bibinfo {title} {Nonlinearity-induced fractional thouless pumping of
			solitons}} (\bibinfo {year} {2025}),\ \Eprint
	{https://arxiv.org/abs/2502.06131} {arXiv:2502.06131} \BibitemShut {NoStop}%
\end{thebibliography}

\begin{thebibliography}{99}
	
\bibitem{SM_eisenbergPRL1998}{H. S. Eisenberg, Y. Silberberg, R. Morandotti, A. R. Boyd, and J. S. Aitchison, 
	Discrete spatial optical solitons in waveguide arrays, 
	\href{https://doi.org/10.1103/PhysRevLett.81.3383}
	{Phys. Rev. Lett. \textbf{81,} 3383 (1998)}.} %%
	
\bibitem{SM_morandotti2001self}{R. Morandotti, H. S. Eisenberg, Y. Silberberg, M. Sorel, and J. S. Aitchison, 
	Self-focusing and defocusing in waveguide arrays, 
	\href{https://doi.org/10.1103/PhysRevLett.86.3296}
	{Phys. Rev. Lett. \textbf{86,} 3296 (2001)}.} %%
	
\bibitem{SM_fleischerNat2003}{J. W. Fleischer, M. Segev, N. K. Efremidis, and D. N. Christodoulides, 
	Observation of two-dimensional discrete solitons in optically induced nonlinear photonic lattices, 
	\href{https://doi.org/10.1038/nature01452}
	{Nature \textbf{422,} 147 (2003)}.} %%
	
\bibitem{SM_mandelik2004gap}{D. Mandelik, R. Morandotti, J. S. Aitchison, and Y. Silberberg, 
	Gap solitons in waveguide arrays, 
	\href{https://doi.org/10.1103/PhysRevLett.92.093904}
	{Phys. Rev. Lett. \textbf{92,} 093904 (2004)}.} %%
	
\bibitem{SM_keLPR2016}{Y. Ke, X. Qin, F. Mei, H. Zhong, Y. S. Kivshar, and C. Lee, 
	Topological phase transitions and {Thouless} pumping of light in photonic waveguide arrays, 
	\href{https://doi.org/10.1002/lpor.201600119}
	{Laser Photon. Rev. \textbf{10,} 995 (2016)}.} %%
	
\bibitem{SM_AAH_1}{P. G. Harper, 
	Single band motion of conduction electrons in a uniform magnetic field, 
	\href{https://doi.org/10.1088/0370-1298/68/10/304}
	{Proc. Phys. Soc. A \textbf{68,} 874 (1955)}.} %%
	
\bibitem{SM_AAH_2}{S. Aubry and G. Andr\'{e}, 
	Analyticity breaking and {Anderson} localization in incommensurate lattices, 
	%\href{https://doi.org/10.1088/0370-1298/68/10/304}
	{Ann. Israel Phys. Soc. \textbf{3,} 18 (1980)}.} %%
	
\bibitem{SM_lang2012edge}{L.-J. Lang, X. Cai, and S. Chen, 
	Edge states and topological phases in one-dimensional optical superlattices, 
	\href{10.1103/PhysRevLett.108.220401}
	{Phys. Rev. Lett. \textbf{108,} 220401 (2012)}.} %%
	
	
\bibitem{SM_krausPRL2012}{Y. E. Kraus, Y. Lahini, Z. Ringel, M. Verbin, and O. Zilberberg, 
	Topological states and adiabatic pumping in quasicrystals, 
	\href{10.1103/PhysRevLett.109.106402}
	{Phys. Rev. Lett. \textbf{109,} 106402 (2012)}.} %%
\end{thebibliography}
\end{document}